\documentclass[10pt, conference,letterpaper]{IEEEtran}
\usepackage{tabularx}
\usepackage{array}
\usepackage{booktabs}
\usepackage{epsfig,endnotes}
 \usepackage[compress, square, comma, numbers]{natbib}
\usepackage[ruled, lined, linesnumbered, commentsnumbered, longend]{algorithm2e}
\usepackage{multirow}
\usepackage{url}
\usepackage{subfig}
\usepackage{amsmath}
  \usepackage{wrapfig}
\newcommand{\DELF}[1]{\iffalse #1 \fi}

\newcommand\ts[1]{{\color{black}#1}}

\newcommand\bl[1]{{\color{black}#1}}

\newcommand\ta[1]{{\color{black}#1}}
\usepackage{amsthm}
\theoremstyle{definition}

\usepackage{color}
\usepackage{amsfonts}
\usepackage{nicefrac}
\newcommand{\DEL}[1]{\iffalse #1 \fi}

\usepackage{mathtools}          

\newcommand{\squishlist}{
\begin{list}{$\bullet$}
  { \setlength{\itemsep}{0pt}
     \setlength{\parsep}{0pt}
     \setlength{\topsep}{0pt}
     \setlength{\partopsep}{0pt}
     \setlength{\leftmargin}{0.7em}
     \setlength{\labelwidth}{0.7em}
     \setlength{\labelsep}{0.2em} } }

\newcommand{\squishlisttwo}{
\begin{list}{$\bullet$}
  { \setlength{\itemsep}{0pt}
     \setlength{\parsep}{0pt}
    \setlength{\topsep}{0pt}
    \setlength{\partopsep}{0pt}
    \setlength{\leftmargin}{2em}
    \setlength{\labelwidth}{1.5em}
    \setlength{\labelsep}{0.5em} } }

\newcommand{\squishlistthree}{
\begin{list}{$\bullet$}
  { \setlength{\itemsep}{0pt}
     \setlength{\parsep}{0pt}
    \setlength{\topsep}{0pt}
    \setlength{\partopsep}{0pt}
    \setlength{\leftmargin}{1em}
    \setlength{\labelwidth}{1.5em}
    \setlength{\labelsep}{0.5em} } }

\newcommand{\squishend}{
  \end{list}  }

\IEEEoverridecommandlockouts\IEEEpubid{\makebox[\columnwidth]{XXX-X-XXXX-XXXX-X/XX/\$31.00 ~\copyright2022 IEEE \hfill} \hspace{\columnsep}\makebox[\columnwidth]{ }}
\begin{document}


\title{Distributed Training for Deep Learning Models On An Edge Computing Network Using Shielded Reinforcement Learning\vspace{-0.1in}}
\author{
\IEEEauthorblockN{Tanmoy Sen and Haiying Shen}
\IEEEauthorblockA{Department of Computer Science, University of Virginia\\ Email: \{ts5xm, hs6ms\}@virginia.edu}
}


\maketitle
%
%

\begin{abstract}
 With the emergence of edge devices along with their local computation advantage over the cloud, distributed deep learning (DL) training on edge nodes becomes promising. In such a method, the cluster head of a cluster of edge nodes schedules all the DL training jobs from the cluster nodes. Using such a centralized scheduling method, the cluster head knows all the loads of the cluster nodes, which can avoid overloading the cluster nodes, but the head itself may become overloaded. 
To handle this problem, we first propose a multi-agent RL (MARL) system 
that enables each edge node to schedule its own jobs using RL. However,
without the coordination between the nodes, action collision may occur, in which multiple nodes may schedule tasks to the same node and make it overloaded. To avoid these problems, we propose a system called Shielded ReinfOrcement learning (RL) based DL training on Edges (SROLE). In SROLE, each edge node schedules its own jobs using multi-agent RL. 
The shield deployed in a node checks action collisions and provides alternative actions to avoid the collisions. As the central shield node for the entire cluster may become a bottleneck, we further propose a decentralized shielding method, in which different shields are responsible for different regions in the cluster and they coordinate to avoid action collisions on the region boundaries. 
Our container-based emulation experiments show that \ts{SROLE reduces training time by up to 59\% with 29\% lower median resource utilization and reduces the number of action collisions by up to 48\% compared to multi-agent RL and the centralized RL. Our real device experiments show that SROLE still reduces the training time by up to 53\% with 28\% lower median resource utilization than multi-agent RL and the centralized RL.}
\end{abstract}

\section{Introduction}\vspace{-0.05in}

Edge devices are currently used for various applications in many areas including transportation and healthcare~\cite{boateng2019experience,8117585,9259386,curtis2019healthsense}. These applications often deploy machine learning (ML) frameworks using data collected by the edge devices' sensors. ML models have transformed into more complex and larger Deep Neural Networks (DNNs). These DNNs are memory and computationally expensive in training due to their complexity and size. On the contrary, an edge device usually does not have sufficient memory or computation resources to conduct the entire DNN model training job. Thus, a DNN is usually trained (or updated) in the cloud, then compressed and deployed on the edge nodes for inference. Such cloud-based training can generate significant delays when the network is intermittent (e.g., disaster, network congestion), and cannot provide data privacy protection~\cite{8486403} for sensitive applications (e.g., medical records) as the data needs to be transferred to the cloud.
In this case, distributing the job of training or updating a DNN model among only edge nodes becomes a promising solution. 


Recent works~\cite{234813,8486403,Keith} for distributed training on edges handle either data parallelism or model parallelism while involving the cloud at a certain stage. Data parallelism methods~\cite{8486403,Keith} deploy replicas of an entire neural network on the edges, and these edges have their subsets of training data. Each edge processes its training data subset and synchronizes model parameters in a parameter server running on an initiator edge or the cloud.
\bl{Model parallelism methods~\cite{234813} divide a neural network and distribute the shallow (earlier) layers to edges and the deep layers to the cloud and let edges communicate with cloud for model parameters transfer from the previous layer transfer to the next layer.} The data parallelism methods sometimes may not be feasible as deploying a large DNN model on a single edge may not be feasible. In contrast, the model parallelism methods suffer from the long delay of communication and intermittent network between edges and cloud\looseness=-1.
A concurrent data and model parallelism based deep learning (DL) system can handle these problems. 
In such a system~\cite{tc}, clusters of edges are created according to geographical locations, and each cluster trains a replica of the DL model using model parallelism based on its locally collected data. Each cluster has a cluster head that has relatively high capacity, and it assigns the partitions of a DNN model (i.e., tasks) to the cluster edge nodes with a goal of minimizing training time. Each edge within a cluster collects its own data and sends the data to the first-layer node, which collects the sensed data from all cluster edges as the training data of the model replica in the cluster. 
The cluster head assigns tasks to the edges based on the resource demands of the tasks and the available resources of the edges. 
Thus, the cluster head needs to continuously observe the workload conditions of all the edge nodes in its cluster. With this cluster-wide knowledge, the cluster head can avoid overloading the edges in assigning the tasks. However, such a centralized scheduling method imposes a significant workload on the cluster head, 
which ultimately impacts the performance of the training. Recently, RL~\cite{mao,zhu2021network} has also been used for such scheduling in the recent times with similar high load. To lessen this load, we first propose a multi-agent RL method (MARL) that enables each edge node to schedule its own jobs among its neighboring edge nodes (i.e., edge nodes in its transmission range) using RL. 
In MARL, each edge device works as an independent agent and makes the scheduling decision among its nearby edges, thus relieving the cluster head from the extra burden. 
\DEL{Each edge device first observes the states of the resource demand of a layer and availability of itself and its nearby edge devices. Upon observing the resource state, each edge device makes the assignment decide whether this particular edge should run this specific layer or another nearby edge device. Finally, this particular layer is assigned only on one edge device-based on individual edge devices' consensus.}However, without the coordination between the edge nodes, action collision may occur, in which multiple nodes may schedule tasks to the same node and make it overloaded.

\DEL{in multi-agent RL different edges can individually decide to run multiple layers on one device, which leads to the overloading of that device, and thus slowing down the training. Consequently, we need the shielding~\cite{ElsayedAly2021SafeMR} to avoid such overloading. We define the overloading of an edge device-based on its utilization of any resource type (memory, cpu and bandwidth). That is, as long as one resource in a device is overloaded, the device is considered as overloaded. The utilization of a type of resource on an edge is defined using the ratio ($D_t(d) / C_t(d)$), where $D_t(d)$ denotes the cumulative resource demand of type $t$ from all the assigned layers to edge device $d$ and $C_t(d)$ denotes  the total capacity of type $t$ in $d$. If the assignment of a specific layer on edge makes this ratio higher than a threshold for one resource, we consider the device will be overloaded.}

To \bl{avoid this problem,} we propose Shielded ReinfOrcement learning based DL Training on Edges (SROLE) on top of the MARL-based assignment approach. The shielding approach~\cite{ElsayedAly2021SafeMR} works as a separate monitor that suggests alternative actions to avoid action collision by observing the states and actions that will be taken by the agents. 
In SROLE, each edge node schedules its own jobs using MARL, 
and the shield, which is deployed on the cluster head, checks the action collisions among the schedules of the edges in its cluster. Edge nodes report to the shield their action decisions, and it checks action collisions and provides alternative actions to avoid the collisions. However, the computational cost of a centralized shield grows dramatically with the number of edges in a cluster, and it may become a bottleneck for the entire cluster. Thus, we further propose a decentralized shielding method, in which different shields are responsible for different regions in the cluster and they coordinate to avoid action collisions on the region boundaries. Specifically, 
a large cluster is divided into multiple sub-clusters according to the geographical proximity, and a shield monitors the edges within each sub-cluster and communicates with its neighboring shields for the edges at the boundary of the sub-clusters to avoid action collisions, i.e., unsafe actions.  The computational cost of each shield in the decentralized method is lower than that in the centralized shielding as the centralized shield's workload is distributed to a number of shields. 

\DEL{In our case, the shield is responsible for observing all edges' final decisions along with the change of available resource state. Upon observing the decision and change of available resource state the shield corrects any decision that causes overloading of the devices \bl{within a cluster}. The shield is deployed on the edge with the highest resource availability at the start of the training. When the shield observes the final decision causes overloading of one device, it suggests an alternative device to avoid overloading. 
This extra load of shielding is not significant compared to that of cluster head-based centralized scheduling because it involves only the cases when the decision is unsafe made-based on already made decisions of individual edges. However, the computational cost of centralized shield grows exponentially with the number of edges in a cluster.}

In summary, the contributions of this work are as follows:
\squishlist
\item To avoid overloading a cluster head due to scheduling DL training jobs in its cluster, we initially propose a multi-agent RL-based method (MARL) that enables each edge node to use RL to schedule its own jobs. For a given DL training job, an edge node makes scheduling decisions for DNN partitions among its nearby edges depending on their resource availability to minimize training time. 

\item To avoid action collisions in MARL, we use the shielding approach in each cluster. The shield in a cluster collects the scheduling decisions of all edge nodes in the cluster, checks the action collisions and provides alternative actions to avoid the action collisions.

\item To avoid overloading a shield in a cluster, we distribute the shielding workload to multiple shields in a cluster and each shield is responsible for a sub-cluster. The neighboring shields communicate with each other to avoid action collisions from the edges on the boundaries of sub-clusters.


\item We measured the performance of the SROLE system on container based emulation on Amazon EC2 instances. Our evaluation shows that the SROLE system shows up to 59\% reduction in training time and up to 48\% reduction in the number of action collisions compared to the centralized RL and MARL approach. \ts{Our real device experiments also show up to 53\% reduction in training time and up to 46\% reduction in the number of action collisions in comparison with the centralized RL and MARL approach.}
\squishend

 The rest of the paper is organized as follows. Section \ref{sec:related-work} presents the related work. Section \ref{sec:design} describes our SROLE system. Section \ref{sec:evaluation} presents the performance evaluation. Finally, Section \ref{sec:concl} concludes the paper with remarks on our future work.

\section{Related Work}
\label{sec:related-work}\vspace{-0.05in}

Researchers have been studying federated ML training and DNN  partition distribution across cloud/fog and edge devices~\cite{234813,Keith,46,136,137,149,151,Wang2018}. In federated learning~\cite{Keith}, edge devices update a trained model on the cloud. Individual edges download a replica of the model and update the models using their own available datasets. Finally, the updated models from individual edges are aggregated through averaging or using a control theorem on the cloud to produce the final model. Many proposed ML inference approaches partition the ML model distributed edge nodes~\cite{134,135,zhou2019distributing}. The works in \cite{134,zhou2019distributing} simply consider each or multiple convolution layers as a partition, and the work \cite{135} vertically divides the convolutional layers in a convolutional neural network (CNN). The work in~\cite{234813} distributes DNN  partition across cloud/fog and edge devices to accelerate training or inference. The resource-constraint edge devices run lighter (earlier) layers of the model and the cloud or fog run heavier (later) layers of the model. 
Resilinet~\cite{yousefpourresilinet} achieves failure-resilient inference in model-parallel ML at the edge. 
Data parallelism training methods distribute training data among different edges for training and accumulate training updates. Wang \emph{et al.}~\cite{8486403}  analyzed the convergence rate of replicas for ML models such as Support Vector Machine (SVM) and K-means, and accordingly proposed a control method that dynamically adjusts the frequency of global update from each replica to the ML initiator in real time to minimize the learning loss under a fixed computation resource budget for the edges. 


To efficiently run ML models on edge devices in terms of inference time, energy and memory, researchers have introduced techniques for compressing the neural networks (NNs)~\cite{121,122,124,129,Ashok2018n2n,kusupati2018fastgrnn,zhang2018shufflenet,yao2018fastdeepiot,polino2018model,malinin2020ensemble,li2020block,nekrasov2019fast,cheng2018model,saputra2019distilling}. For example, the works in~\cite{deepiot,yang-comp,Liu:2018} find compressed DNN models by formulating optimization problems that meet resource (memory, energy) constraints or minimize inference time while maximizing accuracy or reaching a specified accuracy. 
Han \emph{et al.}~\cite{deep-compression} proposed cumulative pruning of the network connection, weight quantization, and compression through Huffman coding to decrease the size and inference time of an NN model without significant loss of accuracy.
Ashok \emph{et al.}~\cite{Ashok2018n2n} proposed a reinforcement learning based policy that first removes layers from the DNN and later reduces the size of the remaining layers by deleting links. Yao \emph{et al.}~\cite{Yao-deep-iot} used a pre-trained compressor-critic network to estimate the link weights and drop out the low-weight links.  


However, there are few works on scheduling the partitions of a DNN model in model parallelism on the edge in an efficient manner. This paper addresses this problem.

\section{Background}\vspace{-0.05in}

\begin{figure}[h]
  \centering
\includegraphics[width=0.64\linewidth,height=0.137\textheight]{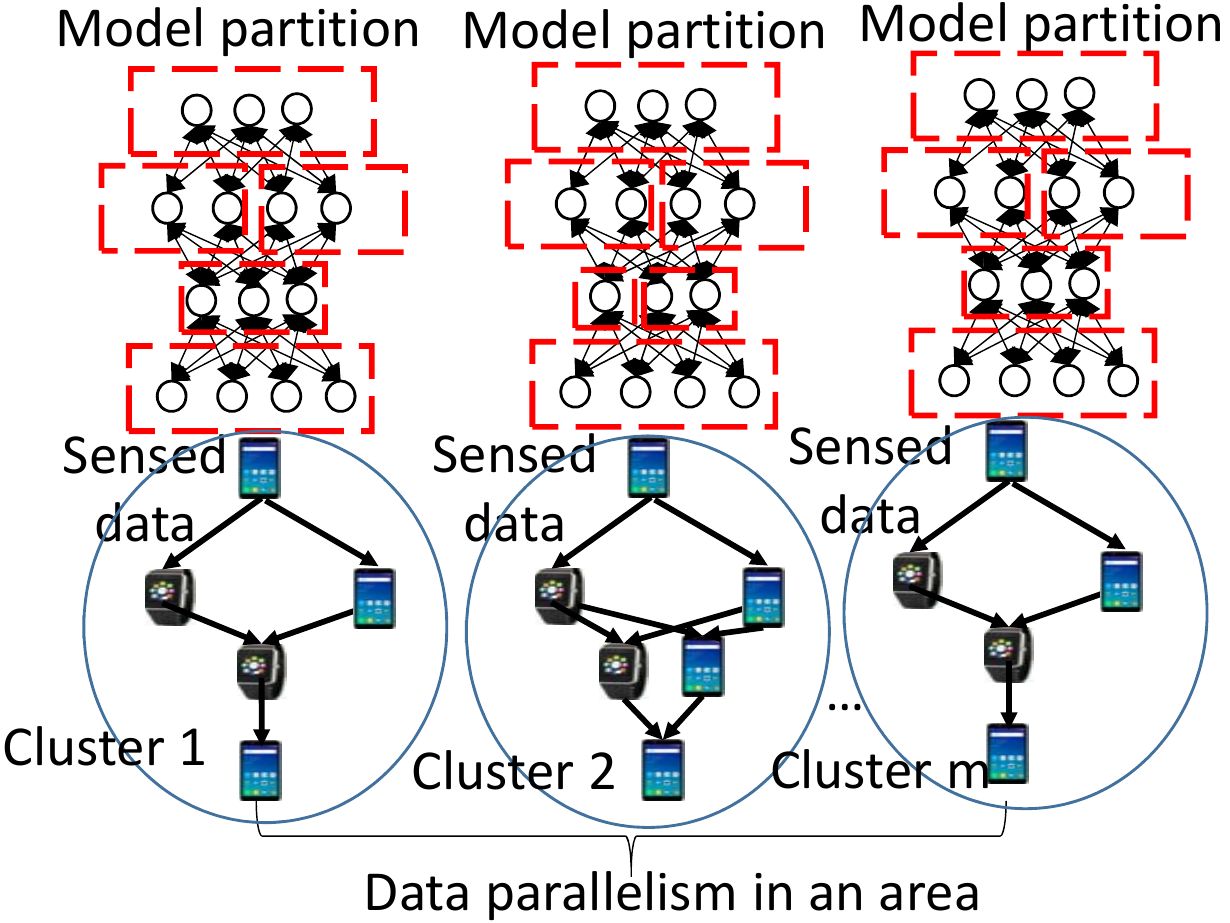}
\caption{Model and data parallel ML on edges.}
\label{fig:Illustration}
\end{figure}

In the model parallelism, a layer is a DNN unit such as convolutional or fully-connected layer~\cite{PipeDream}. In each model level, a model partition consists of one or multiple disjoint layers, which can be executed in parallel. These partitions are assigned to the edge nodes based on their available resources. In this paper, we use concurrent data/model parallelism as an example to explain our proposed methods though they also can be applied to model parallelism. The problem we handle is how to schedule the different model partitions (i.e., tasks) to different edge nodes to minimize the training time. In concurrent data/model parallelism, as shown in Figure~\ref{fig:Illustration}, 
each cluster is formed by proximity-close edge nodes. Each cluster trains an entire DNN model replica using the cluster's collected dataset, and partitions the model among the proximity-close cluster edges. The cluster head is responsible for initiating the training, partitioning and distributing the layers of the model to edge nodes, and synchronizing model parameters to generate the final trained model. 
In order to conduct task scheduling, the cluster head needs to continuously check the resource availability of all the edge nodes in its cluster and also estimate the resource demands of each partition in scheduling the tasks of a job. The resource availability and resource demands are for multiple resources, mainly including GPU or CPU, memory, and bandwidth.

An edge node is considered overloaded when the sum of the resource demands of its running tasks is larger than its resource capacity for one type of resource. 
If an edge node running a model partition becomes overloaded, the training process may slow down. Thus, each device $d_j$ measures its resource utilization of each type type-$k$ resource periodically at each timestep $t$ as follows: 
\begin{equation}
\label{eq:overload}
\small
u_k(d_j) = \frac{D_k(d_j)}{C_k(d_j)}, 
\end{equation} where $D_k(d_j)$ denotes the total resource demand of type-$k$ resource of the tasks running on edge $d_j$ and $C_k(d_j)$ denotes the capacity of type-$k$ resource of edge $d_j$. 
The system pre-defines $\alpha$ (e.g., 0.95) and if $u_k(d_j)>\alpha$ for any type-$k$ resource, the edge node is considered as overloaded. We define an edge node's combined resource utilization as follows:
\begin{equation}
\label{eq:overload}
\small
u(d_j)=\prod u_k(d_j), ~k=1, 2, ...
\end{equation} It measures the overall resource utilization across different types of resources of an edge node.


The cluster head can use RL for the task scheduling and functions as the agent in the RL. RL has three components: state, action and reward. Given the current state, the agent chooses the action that generates the maximum expected reward and receives reward for the action it takes. In this task scheduling scenario, the state is the resource demand of each layer in the DL model and the resource availability of each edge node in the cluster. The action is the schedule that assigns each partition to an edge node. The reward is defined based on the training time of the DNN model; shorter training time leads to higher reward and vice versa. Thus, using the trained RL, the cluster head observes the state and makes the scheduling decision for each DL training job. However, all of these operations create significant overhead on the cluster head. In this paper, we aim to distribute the overhead among the cluster edge nodes while decreasing training time increase due to this decentralized operation.



\section{System Design of SROLE}
\label{sec:design}
\vspace{-0.05in}
\subsection{Overview}
\vspace{-0.05in}
We propose SROLE that consists of the following components.

\begin{itemize}

\item \textbf{Multi-agent RL (Section~\ref{sec:MARL}).} To distribute the job scheduling overhead on the cluster head among the cluster edge nodes, we
propose a multi-agent RL-based method (MARL). In MARL, each edge node uses RL to schedule the tasks of its own DL training job without relying on the cluster head.

\item \textbf{Centralized Shielding for MARL (Section~\ref{sec:cShielding}).} Edge nodes share their neighbors since there are overlaps in the transmission ranges of neighboring edge nodes. Then, edge nodes may schedule their tasks to the same edge node since they do not know the decisions of other edge nodes, which may overload the task assigned node. Thus, we propose a centralized shielding method for MARL, which checks the action collisions and provides alternative actions to avoid the action collisions in a cluster.

\item \textbf{Decentralized Shielding for MARL (Section~\ref{sec:dShielding}).} As the centralized shield is deployed in one edge node, which may overload it, we propose a decentralized shielding method for MARL. In this method, multiple shields are deployed to the sub-clusters in one cluster and each shield is responsible for its own sub-cluster. Further, the neighboring shields communicate with each other to avoid the action collisions on the boundaries.

\end{itemize}

\subsection{Multi-Agent RL-based Job Scheduling}\label{sec:MARL}
\vspace{-0.05in}
The MARL method is similar to the above RL-based method, except each edge node is an agent and the state includes the resource availability of an edge node's nearby nodes rather than all the edge nodes in the cluster. The RL is initially pre-trained and distributed to each edge node. As a result, each edge node uses the RL to schedule the tasks of its own DL training jobs and keeps training the RL model.
In this method, each edge takes the optimal action of assignment of the partitions based on its observed state.
In particular, each edge makes its own local decision on where each layer should be assigned. 
\begin{wrapfigure}{r}{0.53\columnwidth}\vspace{-0.05in}
\centering
\hspace{-0.15in}
   \includegraphics[width=0.55\columnwidth, height=0.137\textheight]{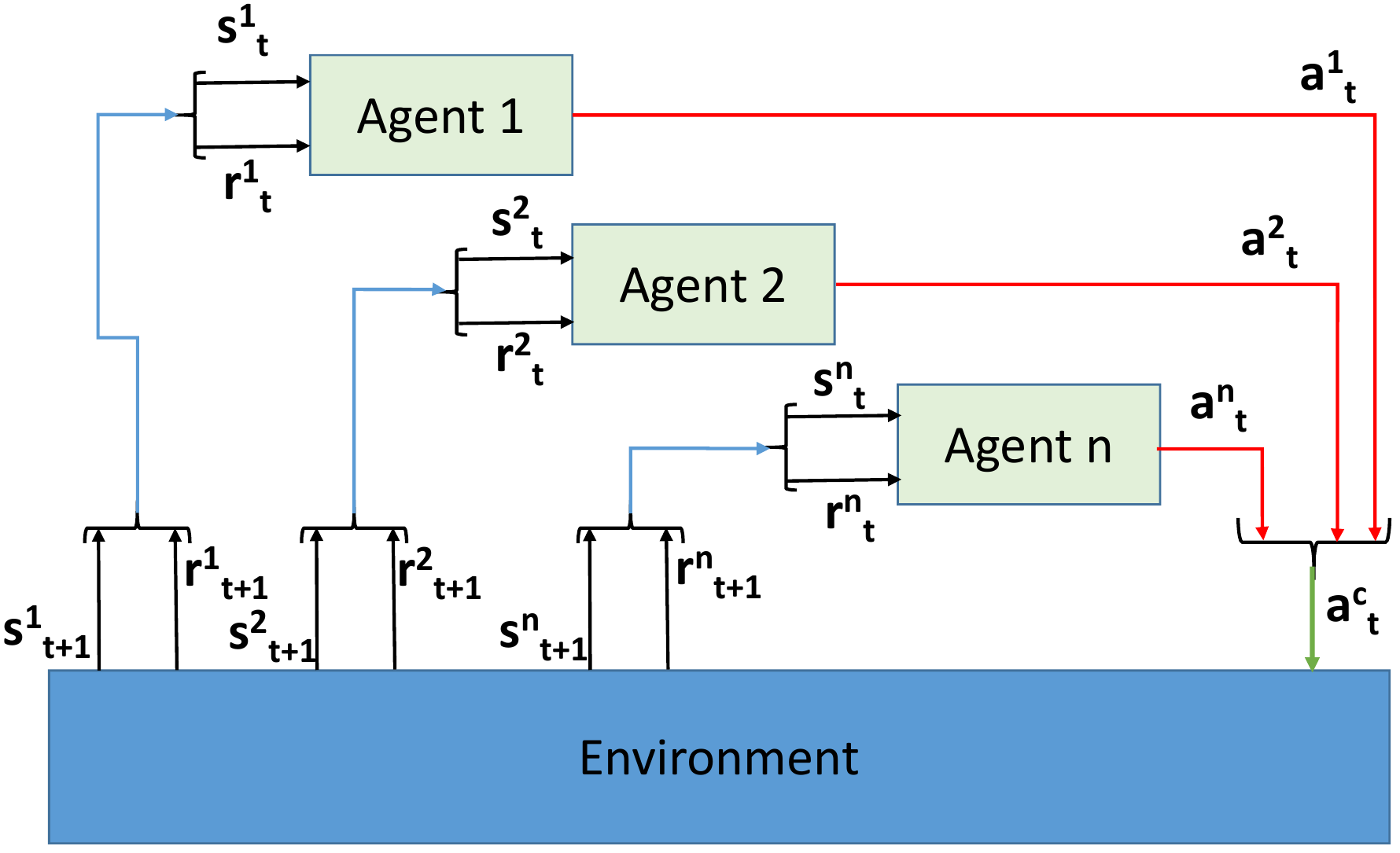}\vspace{-0.15in}
   \vspace{-0.05in}\caption{The process of multi-agent RL.}\vspace{-0.1in}
   \label{fig:rl}\vspace{-0.1in}
\end{wrapfigure}
Based on the decision made by each edge node, the state of the environment changes as the available resources change for the edge node where a layer is assigned. Consequently, based on the local decision taken by one edge or agent, the global decision of all agents influences the overall state. Finally, each edge node has its own long-term reward to optimize, which now becomes a function of the policies of all other agents that are updated based on the global decision. Figure~\ref{fig:rl} illustrates the working procedure of the multi-agent RL. Each edge node observes the state space from the environment and then takes its own action to assign partitions to itself and its neighbors, and then it receives reward. The joint action of the actions of all agents are denoted by $\mathbf{a^c_t}$: $\mathbf{a^c_t}=\mathbf{a^1_t}\bigcup \mathbf{a^2_t} ... \mathbf{a^i_t}...\bigcup \mathbf{a^n_t}$. After the actions are taken, the state $\mathbf{s}$ and reward $\mathbf{r}$ at the next timestep $t+1$ (i.e., $\mathbf{s_{t+1}}$ and $\mathbf{r_{t+1}}$) become the state and reward at this timestep $t$ (as indicated with arrows) for the agents to make decisions again. 


Now, we explain the corresponding state (S), action (A) and reward (R) for our proposed MARL model by each agent.

\noindent{\textbf{State space.}} The state space ($\mathbf{s_t}\in S$) consists of the resource demands of all the layers of a DNN model, and the available resources of all of a node's nearby edge nodes. 
For each layer, the state includes the CPU resource demand, memory demand, and its data transfer size to each layer in the next level in the DNN model. Besides, the state also includes the utilization of the CPU, memory and bandwidth resource for each device.
The state of edges includes the available CPU and memory of each edge, and available bandwidth across each pair of edges at each time $t$. As the continuous values of these resource characteristics result in infinite size of the state space, we discretize the continuous space by dividing their value range into a number (e.g., three) of equal-width ranges: low, medium and high.

We varied the structural parameters of a particular DNN layer structure within reasonable ranges as indicated in~\cite{fastdeepiot} and profile the CPU and memory usage in the forward and the backward pass. We use the TensorFlow benchmark tool~\cite{MLPERF} to profile the usage of all DL components on an edge node. The available resources on an edge device keeps changing accordingly to the layers assigned to the device.

\noindent \textbf{Action space.} We use $\mathbf{a_t}$ to denote $\mathbf{a^i_t}$ for simplicity. The action space represented by $\mathbf{a_t}$$=\{a^{i,j}_t\}\in A,~$($i=1,2,...,|M|,~j=1,2,...,|E|$) defines the
schedule for all the layers at time $t$, where $M$ denotes the set of the layers in the DNN model, $E$ denotes the set of all nearby edges and $|\cdot|$ means the size of a set.
Each element of the action space $A$ defines which edge should be assigned with a certain layer.
Action $a^{i,j}_t$ is defined as follows for each pair of layer $l_i$ and edge device $d_j$. 
\[ a^{i,j}_t =
    \begin{cases}
     1, & \text{if layer $l_i$ is placed in edge $d_j$ at $t$} \vspace{-0.0in}\\
     0, & \text{otherwise}\vspace{-0.0in}
 \end{cases}\vspace{-0.1in}
 \]
 After $\mathbf{a_t}$ is determined, the available resources of edges are updated with the available resources at $t+1$, and then $\mathbf{a_{t+1}}$ is determined, and so on until the last layer is assigned. 

\noindent \textbf{Reward.} 
Let $\mathbf{a}_t$ be the action taken (schedule is made) at time $t$, then the reward function  is given by:\vspace{-0.1in}
\[ {\mathbf{r_t}}(\mathbf{s_t},\mathbf{a_t}) =
\begin{cases}
    -\gamma, & \text{\small{if memory is violated}} \\
    \frac{\rho}{\sqrt{O}},  & \text{otherwise}
\end{cases} \vspace{-0.05in}
\]where $\rho$ is a coefficient to control the reward, $\gamma$ is a large constant reward to ensure that a schedule violating the memory limit requirement is not valid. Furthermore, and $O$ denotes the training time of the DNN model. 
After the job assignment, the states change for the next assignment and the reward is updated for all agents.

\subsection{Centralized Shielding for MARL}\label{sec:cShielding}
\vspace{-0.05in}
The MARL method cannot ensure that none of the edge nodes get overloaded since different edge nodes may assign tasks to the same node simultaneously based on its original available resources. To handle this problem, we propose a shielding approach on top of the multi-agent RL scheme. Each cluster has a shield deployed in the cluster head that has high resource capacity. It ensures that none of the edges get overloaded by the task assignment from all edge nodes in the cluster. 
In the centralized shielding method, the shield enforces the safety specification (i.e., avoiding decisions that overload an edge before being sent to the environment) during the RL learning process. 
After an edge node makes a scheduling decision for its job, it reports its decision to the shield in its cluster. The shield collects the decisions of all edge nodes in its cluster and checks action collisions, i.e., the actions that will make an edge node overloaded by hosting the tasks from multiple edge nodes, and then provides alternative actions to avoid the action collisions.
\begin{wrapfigure}{r}{0.53\columnwidth}\vspace{-0.151in}
\centering
\hspace{-0.15in}
   \includegraphics[width=0.55\columnwidth, height=0.137\textheight]{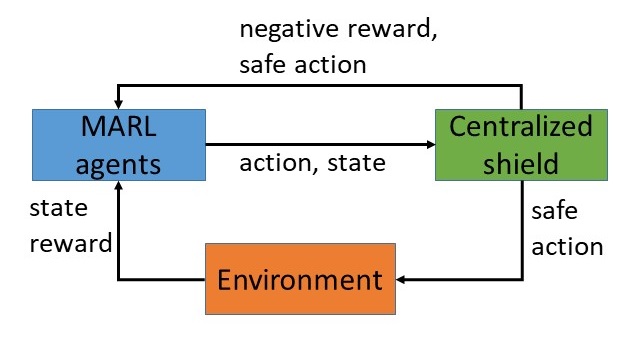}\vspace{-0.15in}
   \vspace{-0.12in}\caption{Centralized shielding.}\vspace{-0.05in}
   \label{fig:crl}\vspace{-0.1in}
\end{wrapfigure}
We hope that the shield restricts MARL agents as less as possible via the minimal interference criteria. These criteria are as follows: (1) the shield only corrects joint action $\mathbf{a^c_t}$ if it violates the safety specification, and (2) the shield seeks a safe joint action $\mathbf{\tilde{a}^c_t}$ that changes as a few of the agents' actions as possible in $\mathbf{a^c_t}$. Figure~\ref{fig:crl} shows the working procedure of the centralized shielding built on top of the multi-agent RL scheme. The centralized shield observes the joint action and current state before the action is implemented. If it leads to an unsafe action, i.e., overloading of any edge, the shield suggests a safe action that will be implemented. At the same time, the shield also notifies the edges within the cluster of the safe action and assigns a constant negative reward ($\kappa$) for their originally decided action that leads to the overload of one device. Accordingly, the reward is redefined as below:

\[ {\mathbf{r_t}}(\mathbf{s_t},\mathbf{a_t}) =
\begin{cases}
    -\gamma, & \text{\small{if memory is violated}} \\
    -\kappa, & \text{suggested by the shield} \\
    \frac{\rho}{\sqrt{O}},  & \text{otherwise}
\end{cases} \vspace{-0.05in}
\]In the meantime, the environment changes based on the suggested safe action by the shield and the states and rewards are updated accordingly.



The shield makes a judgement about the potential violations of safety specifications.
In details, the shield observes whether the joint action actually changes the resource utilization of any type of resource of an edge to a value higher than the threshold, i.e., $u_k(d_j)>\alpha$. 
If this condition is true (criterion (1)), there exists an action collision and the shield will choose alternative safe actions to replace the original actions in the joint action to make the edge node (say $d_j$) not overloaded. 
For each layer $l_i$ that is assigned to edge node $d_j$, we define its resource demand weight as follows:
\begin{equation}\small
\omega(l_i)= \prod_k (b_k(l_i)/C_k(d_j)), ~k=1, 2, ... \vspace{-0.07in}
\end{equation}where $b_k(l_i)$ is the resource demand of type-$k$ resource of layer $l_i$ and $C_k(d_j)$ is the capacity of type-$k$ resource of edge device $d_j$. While choosing the safe action, the shield first ranks the layers that are planned to be assigned to the edge node based on their resource demand weights. Then, it picks up the layer with the highest weight and finds a new host edge for it that will not be overloaded after hosting this layer. The purpose of choosing the layer of the highest weight to be rescheduled is to reduce the interference to the original joint action (criterion (2)). The shield repeats this process until the remaining layers will not overload the edge. Specifically, it searches for nearby edge nodes with high resource availability from edge node $d_j$, and then checks whether any of these edges can host this layer after it accepts other layers that are planned to assign to it in other actions. To quickly find such an edge node, the shield calculates the combined resource utilizations of the nearby edge nodes after they accept other planned layers assigned to them. Next, it orders the nearby edge nodes in the ascending order of their combined resource utilizations and then sequentially picks up the top node to check until it finds such an edge node. The edge node on the top generally has a high available resources and hence is more likely to be able to host the layer. After finding the new host edge, the shield creates an alternative action that assigns the layer to this edge device. As we limit our safe action  from the nearby edges of the original edge node in the decided original action and ensure this newly suggested action won't overload the edge, this new action will not deviate from the previous optimal action greatly. 

\begin{algorithm}
\label{algo}
\DontPrintSemicolon
\small
    \SetKwInOut{Input}{Input}
    \SetKwInOut{Output}{Output}

\For{each timestep}{
Collect actions from all edge nodes in the cluster\;
Virtually take the actions to assign layers to edges\;
    \ForEach{each edge node $d_j$}{
    Calculate resource utilization $u_k(d_j)$ of each resource\;

   Rank the assigned layers on $d_j$ in descending order of resource demand weight\;
   Punishment $\kappa \leftarrow 0$\;
        \While{any $u_k(d_j) > \alpha$, $k=1, 2, ...$}
        {

        Choose the top layer which is in action $\mathbf{a_t}$\;
            $\mathbf{\tilde{a}_t} \leftarrow $ safe action by the shield \;
               Replace $\mathbf{a_t}$ by $\mathbf{\tilde{a}_t}$\;
              $\kappa \leftarrow \kappa +$ constant negative reward\;
              Notify the layer's scheduling edge about $\kappa$ and $\mathbf{\tilde{a}_t}$\;
        }

}

}
    \caption{\small{Pseudocode of the centralized shielding executed by the shield in a cluster.}} \label{algo:kn}
\end{algorithm}

%
%
%

Algorithm~\ref{algo:kn} illustrates the pseudocode for the centralized shielding. At each timestep $t$ (Line 1), the central shield observes the joint action and joint state of all the agents or edges in the cluster (Lines 2-3). 
For each edge node $d_j$, it calculates its resource utilization of each resource type (Line 5), ranks the assigned layers on $d_j$ in descending order of resource demand weight (Line 6), and initializes $\kappa$ (Line 7). It then checks whether its resource utilization is greater than the pre-defined threshold $\alpha$ (Line 8). If yes, the shield picks up the layer on the list top, finds and suggests a safe action for scheduling the layer and notifies the layer scheduling edge about the new action and the $\kappa$ reward for the unsafe action (Lines 9-13). The shield repeats this process until the edge node will not be overloaded after it hosts all layers assigned to it (Lines 8-14). 

\subsection{Decentralized Shielding for MARL}\label{sec:dShielding}\vspace{-0.05in}
A shield in a cluster is responsible for all edge nodes in a cluster and may become overloaded due to the communication and computation overhead in shielding. Thus, we propose a decentralized shielding method. In the
decentralized shielding method, we first divide a cluster to multiple sub-clusters and each sub-cluster consists of geographically proximity-close edge nodes. Then, one shield works for one sub-cluster. Within each sub-cluster, the shield works in the similar way as described in the centralized shielding. There is one additional problem we need to handle. The edge nodes in the boundary of two or more sub-clusters may assign tasks to the same edge node in one sub-cluster, which may overload it, but the shield in this sub-cluster will not receive the actions from its neighboring sub-clusters and hence will not detect the action collision. To solve this problem, the shields of neighboring sub-clusters need to communicate with each other to avoid such a case. Specifically, the neighboring shields first select a delegate to check the action collisions. Then, they send the actions of the edge nodes and the available resources as well as the resource utilizations of the edge nodes in the boundary to the delegate. The delegate uses the same method in the centralized shielding method to check action collision and finds alternative actions. It sends the alternative actions to the neighboring shields, and the shields then forward the alternative actions to the corresponding edge nodes. As a result, the edge nodes take alternative actions instead of previously determined actions.



\section{Performance Evaluation}
\label{sec:evaluation}
\begin{figure*}[!t]
\centering
    \subfloat[VGG-16.\label{fig:low_mem}]{{\includegraphics[width=0.32\linewidth,height=0.137\textheight]{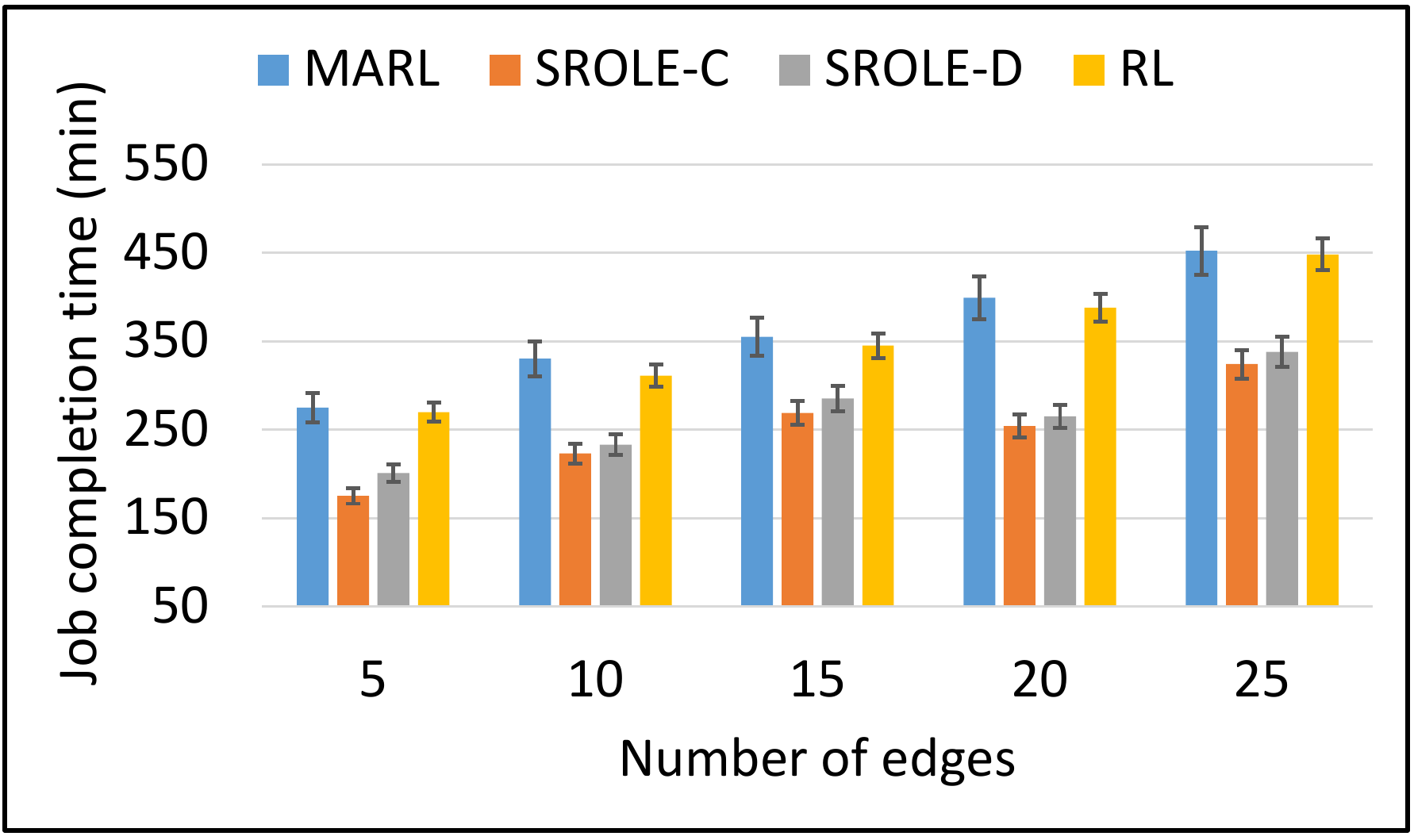} }}%
    \hfill
    \subfloat[GoogleNet. \label{fig:low_cpu}]{{\includegraphics[width=0.32\linewidth,height=0.137\textheight]{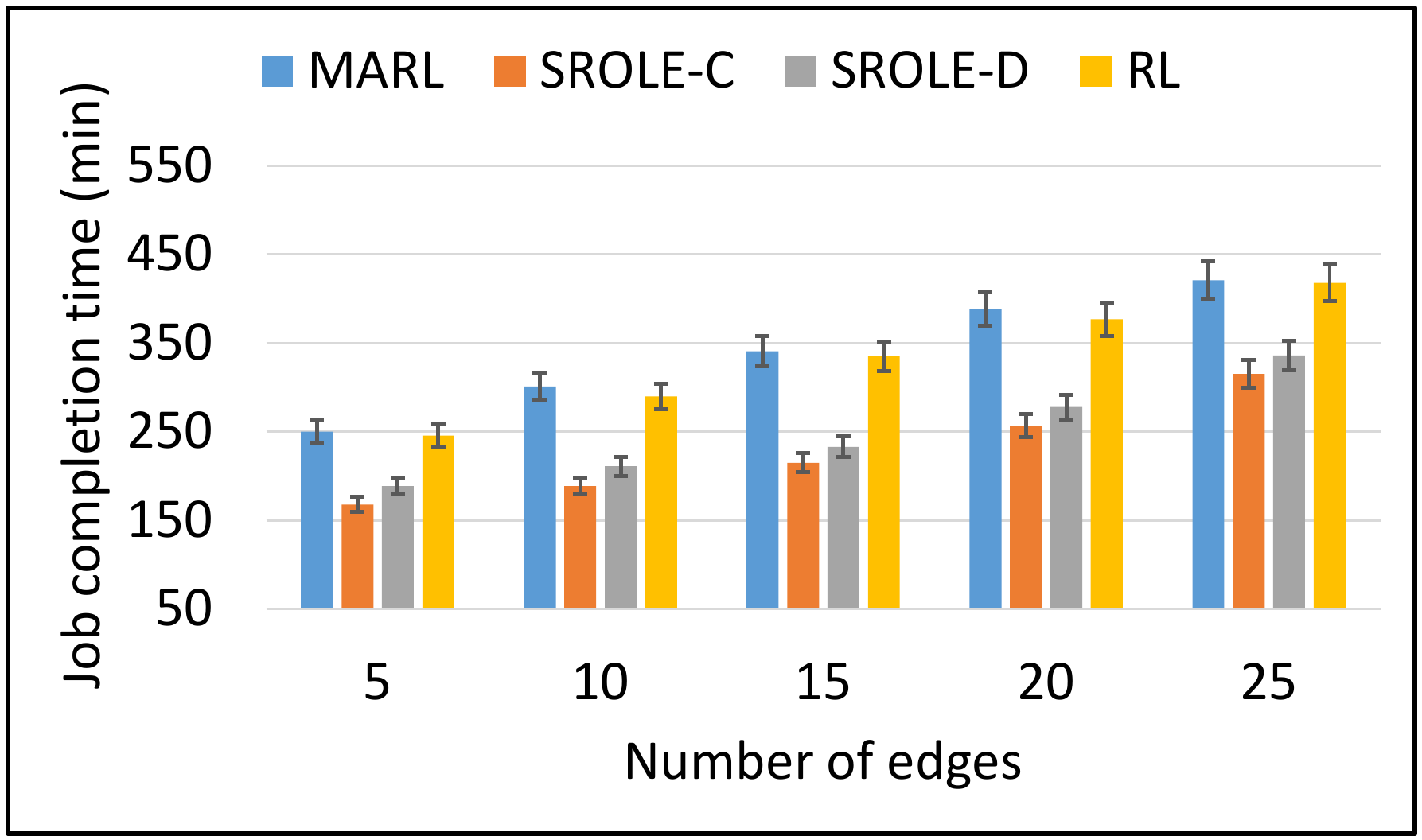} }}%
    \hfill
    \subfloat[RNN.\label{fig:low_bw}]{{\includegraphics[width=0.32\linewidth,height=0.137\textheight]{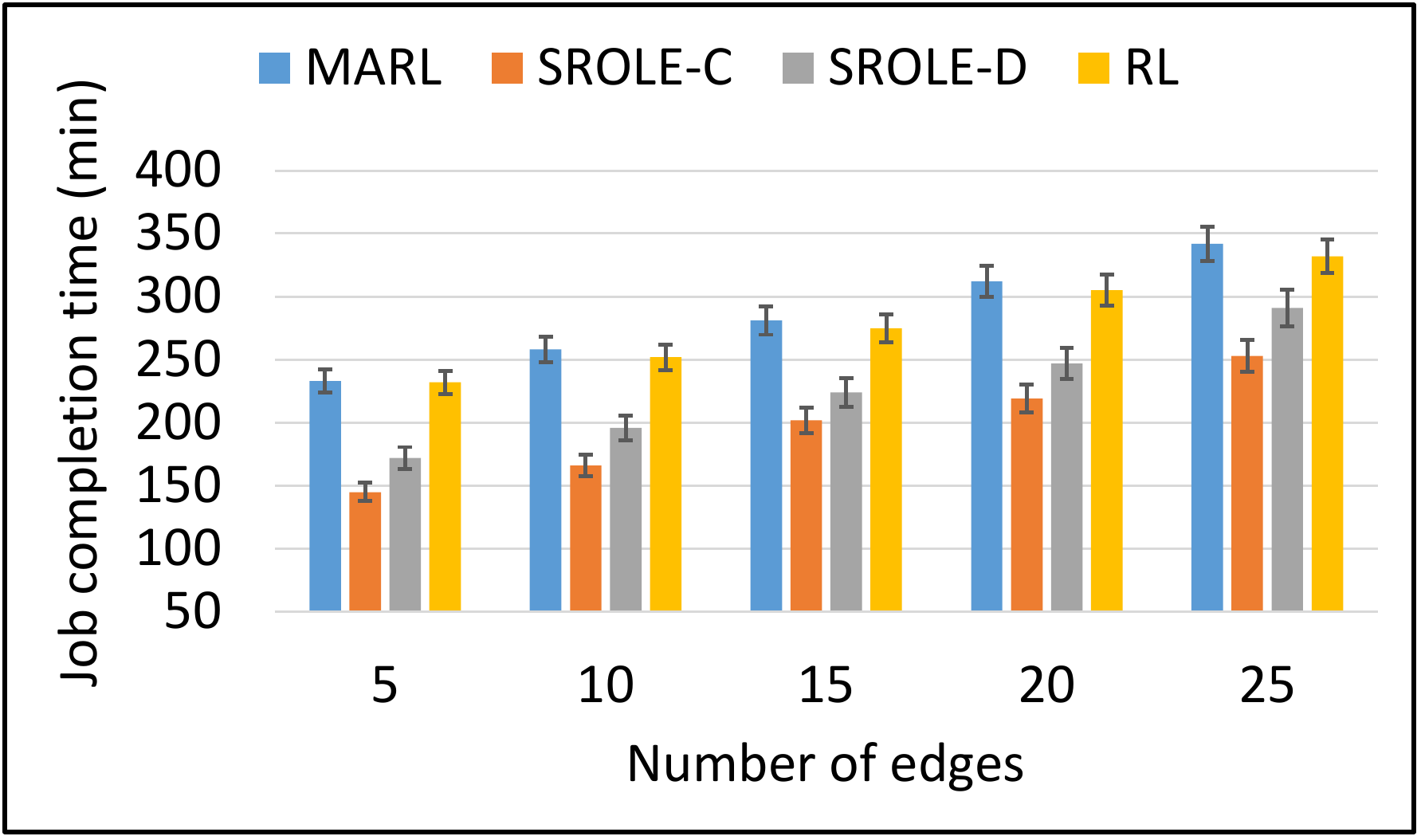} }}%
    \hfill
   \vspace{-0.08in}
   \caption{Job completion time for different models from emulation.}%
    \label{fig:low}\vspace{-0.2in}
\end{figure*}

\begin{figure*}[!t]
\centering
    \subfloat[VGG-16.\label{fig:low_mem1}]{{\includegraphics[width=0.32\linewidth,height=0.137\textheight]{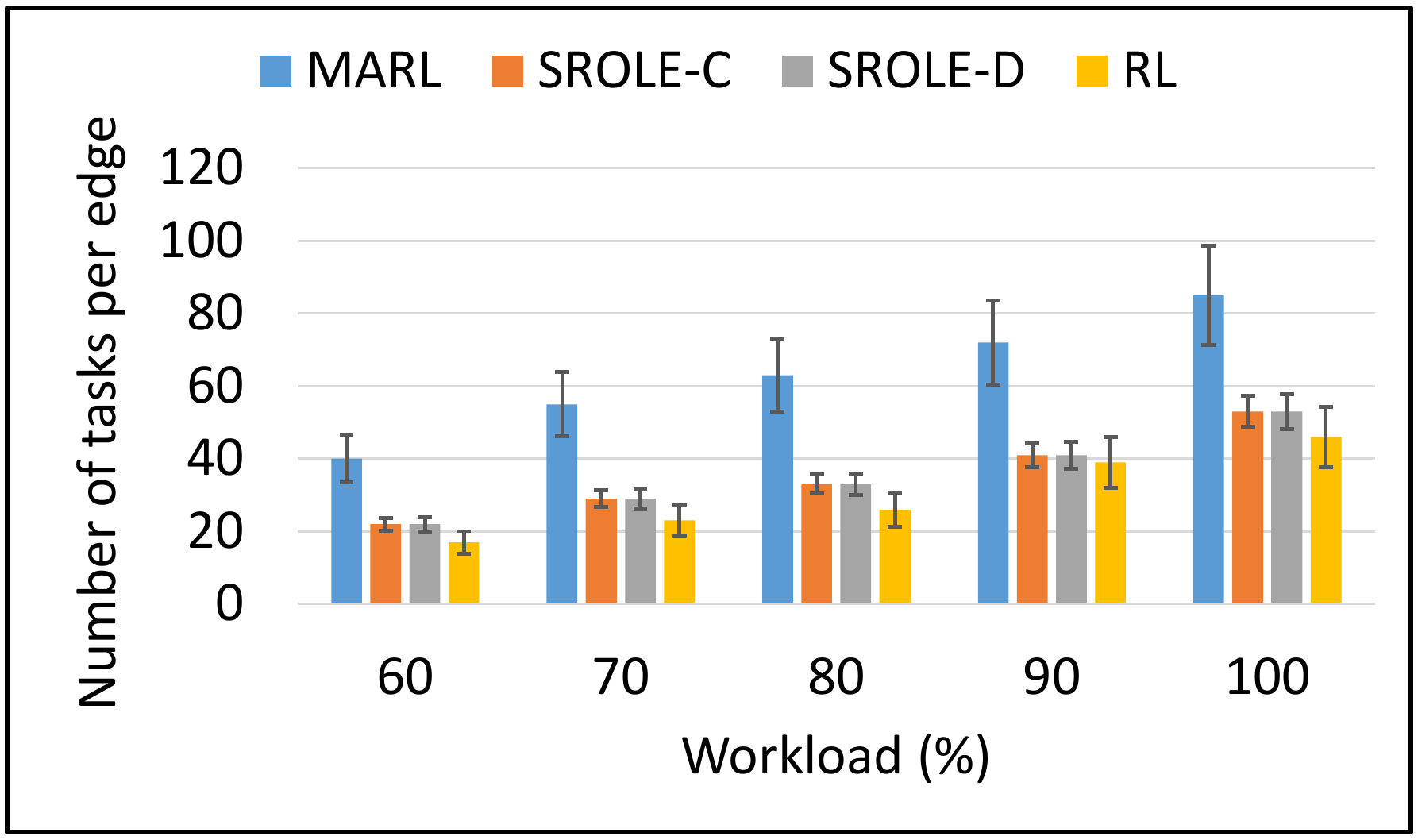} }}%
    \hfill
    \subfloat[GoogleNet. \label{fig:low_cpu1}]{{\includegraphics[width=0.32\linewidth,height=0.137\textheight]{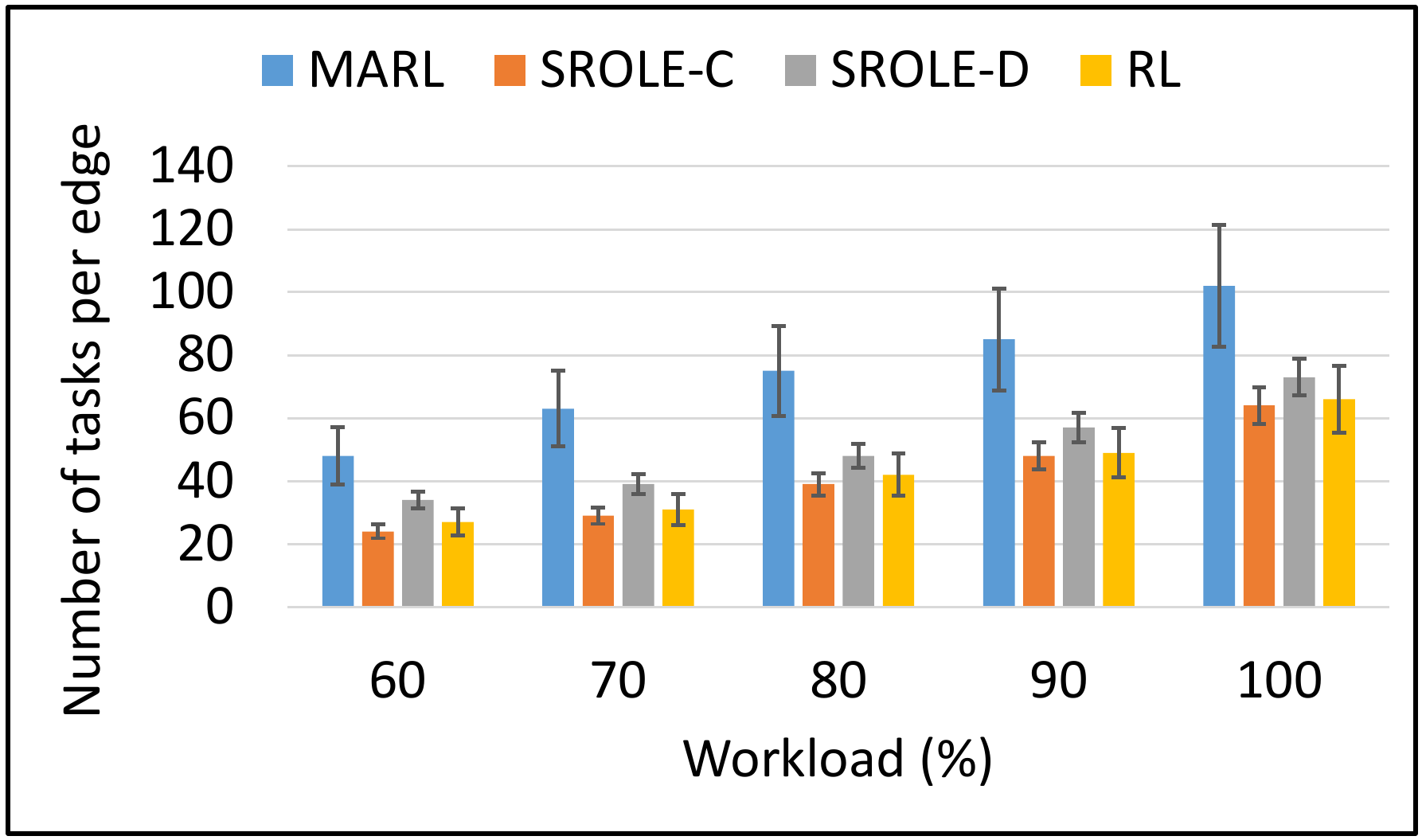} }}%
    \hfill
    \subfloat[RNN.\label{fig:low_bw1}]{{\includegraphics[width=0.32\linewidth,height=0.137\textheight]{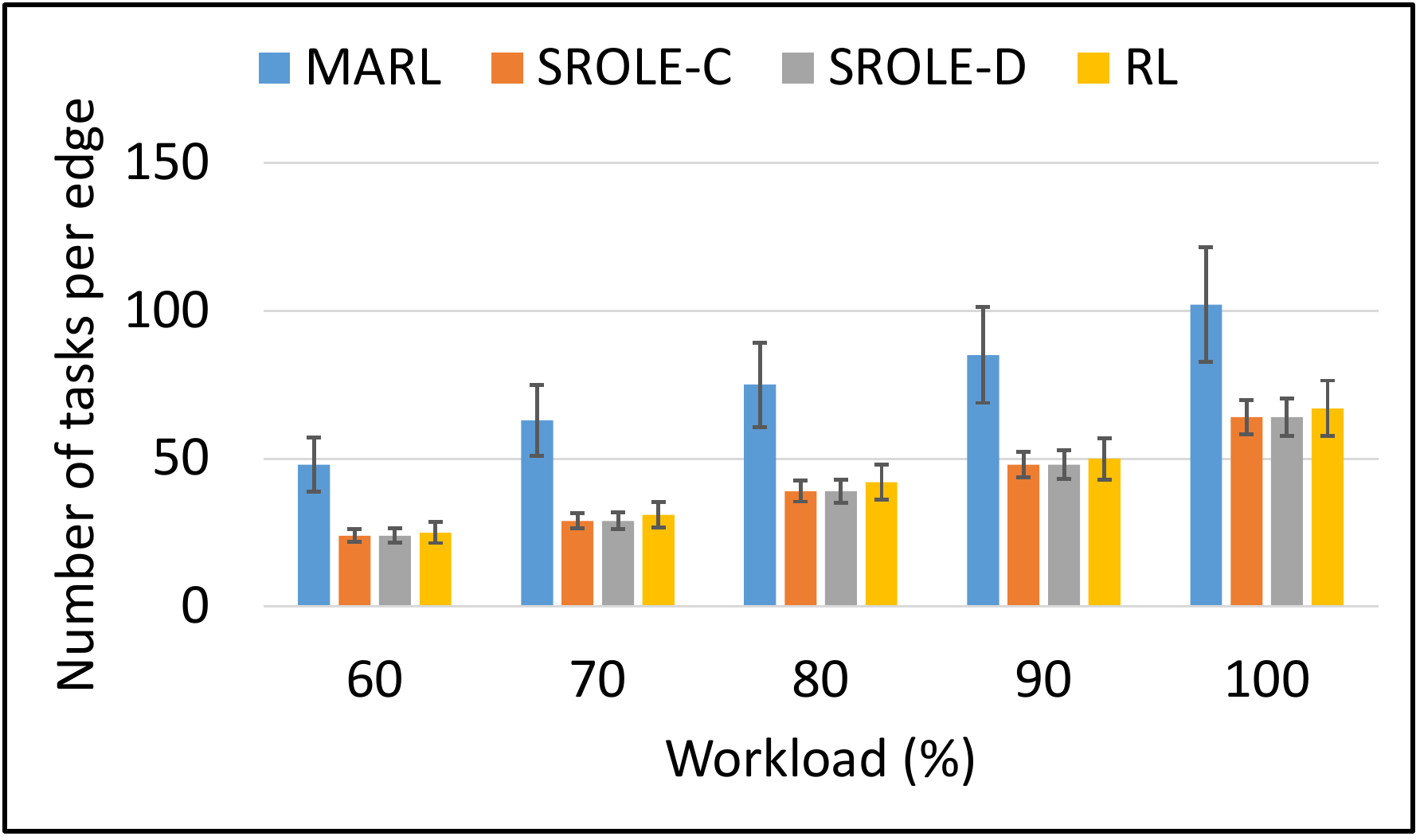} }}%
    \hfill
  \vspace{-0.08in}
   \caption{The number of tasks per device for different models from emulation.}%
    \label{fig:low1}\vspace{-0.2in}
\end{figure*}


%

\subsection{Experiment Setup}
\vspace{-0.05in}
\noindent \textbf{Emulation.} Our proposed system runs on Tensorflow
for the execution of the model training through its parameter
server strategy. In order to emulate edges with varying resources, we use 25 docker containers. Each cluster has 5 edge nodes. The resource settings of our emulation and real device experiments are indicated in Table~\ref{tab:resource_range} and the resources of the devices were assigned in a round-robin way. The containers are deployed in Amazon EC2 instance of type m5ad.4xlarge. The CPU and memory are configured using commands from docker and the bandwidth between different containers is configured using the tcconfig tool~\cite{tcconfig}.

\noindent \textbf{Real experiments.} Our real testbed consists of 10 Raspberry Pis; two Pis have 1 GB memory, four other Pis have 2 GB memory and four other Pi
has 4 GB memory. They roughly have same CPU but we use the cpulimit~\cite{cpulimit} command to control the CPU as desired according to Table~\ref{tab:resource_range}. The edges are connected via 2.4 GHz band wireless connection. We use wondershaper~\cite{wondershaper} tool to control bandwidth among the edges.\looseness=-1

\begin{table}[]
\small
\caption{Resource configuration.}
\begin{tabular}{|p{.25\columnwidth}|p{.65\columnwidth}|}
\hline
          Environment & Resource ranges    \\ \hline
          Real edge & Mem$\in \{1024, 2048, 4096\}$MB \\ & CPU$\in \{0.25,0.5,1.0\}$Host~Ratio \\ & BW$\in \{20,100\}$MBps    \\ \hline
          Container & Mem$\in \{768, 1024, 1536, 2048, 4096\}$MB \\ & CPU$\in [0.3, 1.0]$Host~Ratio \\ & BW$\in \{50, 100, 200, 500, 1000\}$Mbps    \\ \hline

\end{tabular}
\label{tab:resource_range}
\end{table}

\noindent \textbf{ML models and datasets.} We run three ML models: GoogleNet Inception, VGG-16, and RNN~\cite{rnn_blog}. We use the MNIST~\cite{kerascnn} dataset to run the first two models and the Air Quality dataset~\cite{air-quality} for the RNN model. The MNIST dataset consists of 70,000 images of handwritten digits and is widely used for training CNN models. 
The Air Quality dataset contains 9358 instances of hourly averaged responses from an array of 5 metal oxide chemical sensors. One instance refers to the sensor values (as the ML inputs) and the AQI value (as the ML output).
Since there are maximally 5 clusters, We divide the dataset to 5 subsets. Each cluster has a subset as the input training data and the data is randomly distributed among the edges in the cluster as their sensed data. We run three DL training jobs of the same type in each cluster initiated by randomly chosen edge nodes.

\noindent{\textbf{RL Training.}} To train the RL models, we need data related to both DNN models and edge nodes. To generate the data related different DNN model structures and  we profiled and obtained their resource demands. To generate edge node configuration data, we consider the number of edge nodes in the range of [2,10]. For each edge node, CPU is chosen randomly from range [0.5,2] GHz, memory is randomly chosen from range [64,4096] MB~\cite{Fan2017DeadlineAwareTS} and the bandwidth across pair of edge nodes is randomly chosen from range [128,1000] MBps~\cite{RILTA}. Using these data, we train the RL model offline.

\noindent \textbf{Workload and Settings.} In all the cases, we trained one DNN model in each cluster and add several other non-ML jobs (PageRank~\cite{5452747}) from the HiBench benchmark to vary available resources on the edges.  The workloads were controlled by running multiple PageRank job on these edges in a distributed way. We run x=2,3,...,6 PageRank jobs in each cluster
throughout the whole training period to control the workload. Workload of 100\% means there are 6 PageRank jobs running
simultaneously in the system, and other workloads are defined similarly in the decreasing order, i.e., 5 is 90\%, 4 is 80\%, and so on. These jobs were run simultaneously with DNN training jobs until the run completes. During these experiments, we either change the number of edge devices or the workload along the x-axis.  Unless otherwise indicated, the number of edge nodes is 25 and the workload is 100\%. Each run for DNN model experiment was executed for 50 iterations. We repeated each experiment 5 times and plotted the median with the 5th and 95th percentile error bars. \ts{During the experiment, we measured the resource utilization of the devices every 10 minutes. In both the cases of emulation and real device, we set the value of the parameters $\alpha = 0.9$, $\rho =1$, $\gamma = 50$ and $\kappa = -100$.}



\subsection{Compared Methods}\vspace{-0.05in}
\noindent \textbf{MARL.} This is a simple multi-agent RL-based method CQ-learning~\cite{Hauwere} without shielding. 
In particular, both the evolution of the system state and the reward received by each agent are influenced by the joint actions of all agents. That is, each agent has its own long-term reward to optimize, which now becomes a function of the policies of all other agents. 

\noindent \textbf{SROLE-C.} This is a multi-agent RL method with an extra centralized shield. In the centralized shielding, there is a single shield to monitor all agents' joint actions and correct any unsafe action if necessary. That is, the shield observes all the agents and prevents any edge node from being overloaded.

\noindent \textbf{SROLE-D.} This is an extension of the centralized shielding. It has multiple shields on multiple sub-clusters in a cluster and 
each shield is only responsible for the agents in its sub-cluster or a subset of agents in its cluster.

\noindent \textbf{Centralized RL.} In the following figures, we use \emph{RL} for simplicity. This is an RL scheme where the cluster head makes the assignment decision for all the jobs in its cluster. In this method, we assign a negative reward for overloading the memory of a certain device and otherwise, the reward is based on the job completion time. 

\subsection{Metrics}\vspace{-0.05in}
\noindent{\textbf{Job completion time.}} This is the training time, which denotes the time period from the time when a job starts to run after scheduling to the time when the training of the whole model completes. This time period contains multiple number of iterations depending on the size of the dataset. In our case, the training for all the models comprises of 50 iterations. 

\noindent{\textbf{The number of tasks per device.}} From different runs, we measure the number of partitions for a DNN training job and tasks for non-ML jobs running on each device. This measurement is to show the performance of avoiding overloading edge nodes.\looseness=-1

\noindent{\textbf{Computation time overhead.}} Computation overhead refers to the decision making time of each method. It is the time period from the time when a job is initiated to the time when the task assignment schedule of the job is made.

\noindent{\textbf{The number of action collisions.}} We measure the number of action collisions for the negative reward ($\kappa$) for unsafe actions.\looseness=-1


\vspace{-0.05in}
\subsection{Experimental Results from Emulation}


\begin{figure*}[!t]
\centering
    \subfloat[VGG-16.\label{fig:low_mem4}]{{\includegraphics[width=0.32\linewidth,height=0.137\textheight]{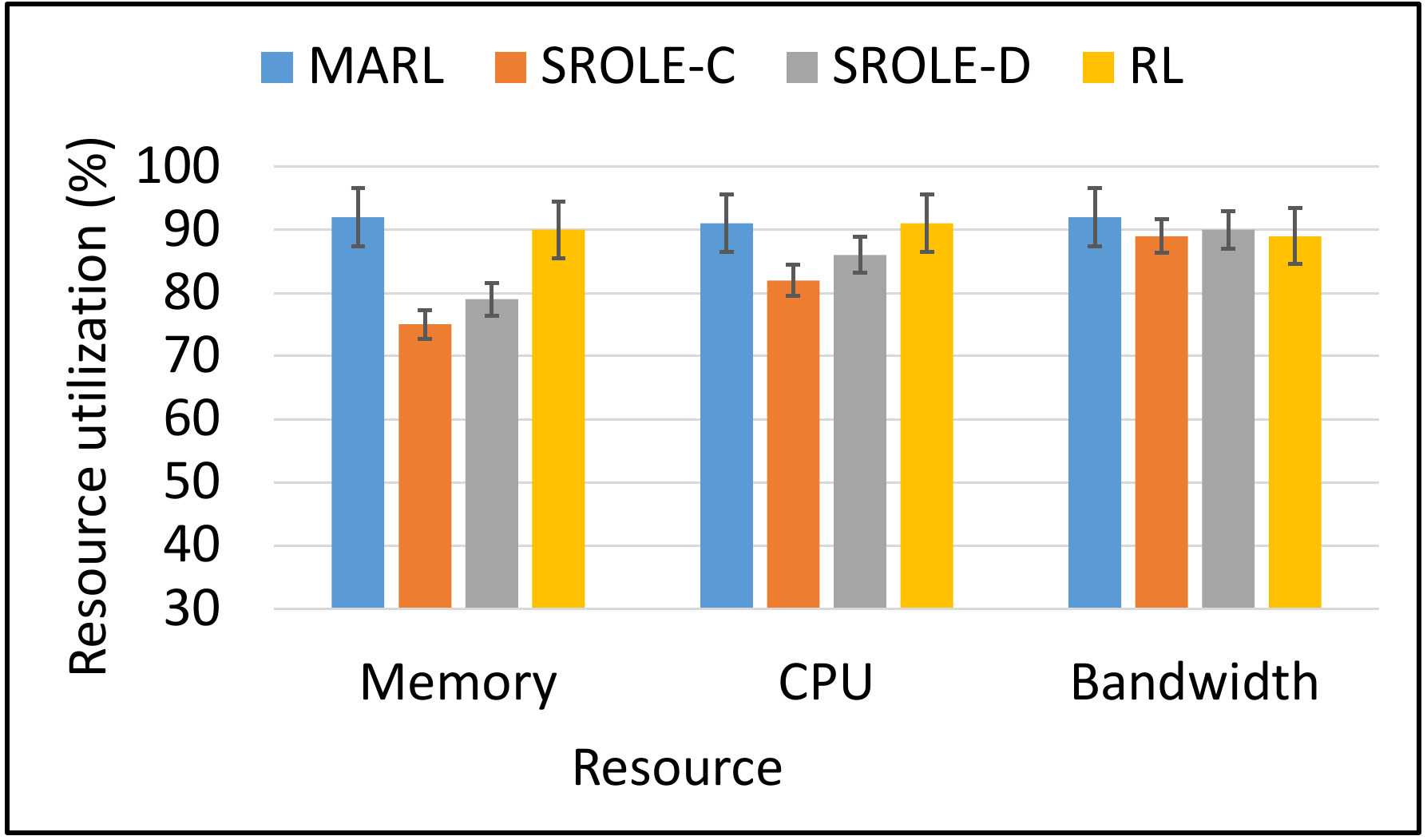} }}%
    \hfill
    \subfloat[GoogleNet. \label{fig:low_cpu4}]{{\includegraphics[width=0.32\linewidth,height=0.137\textheight]{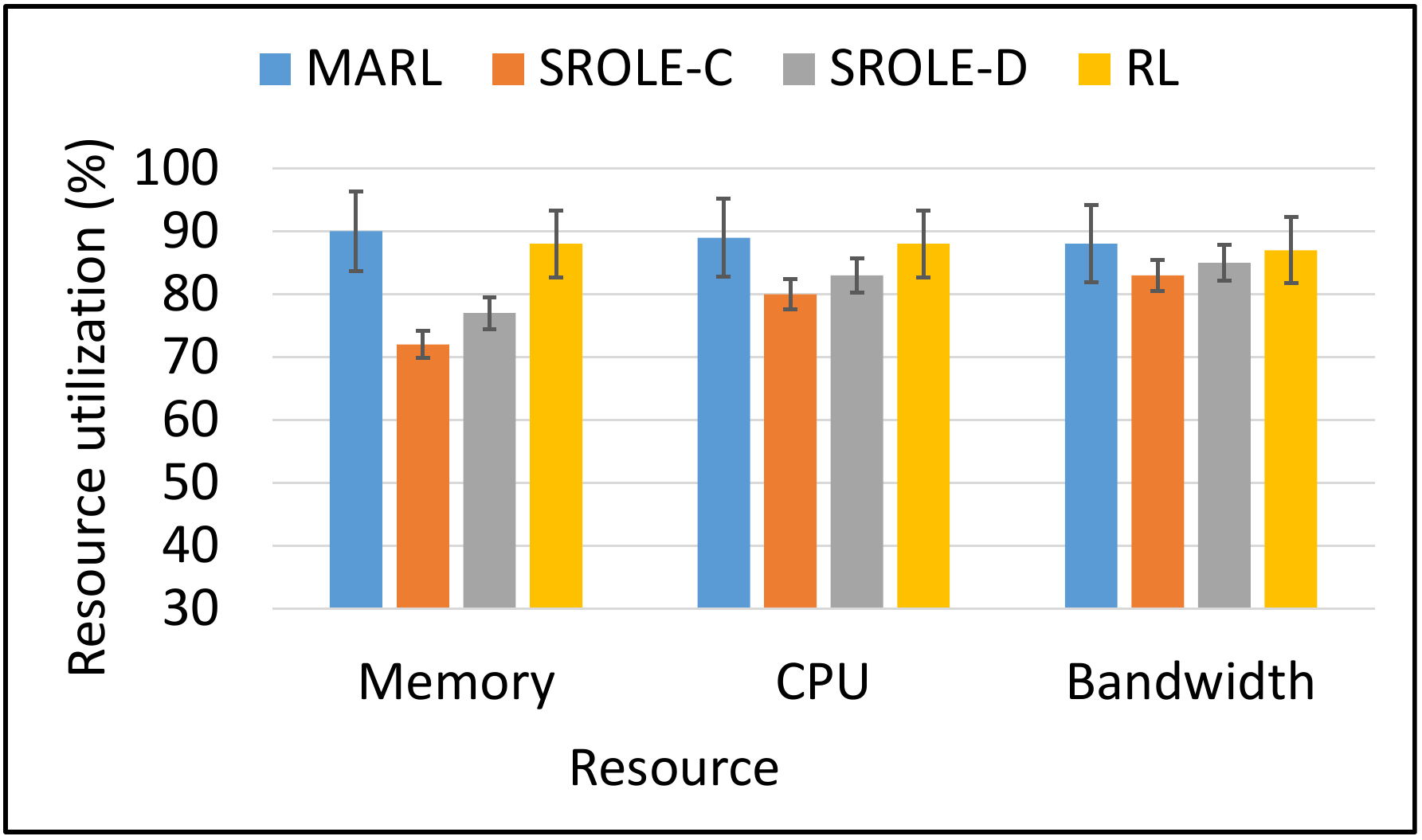} }}%
    \hfill
    \subfloat[RNN.\label{fig:low_bw4}]{{\includegraphics[width=0.32\linewidth,height=0.137\textheight]{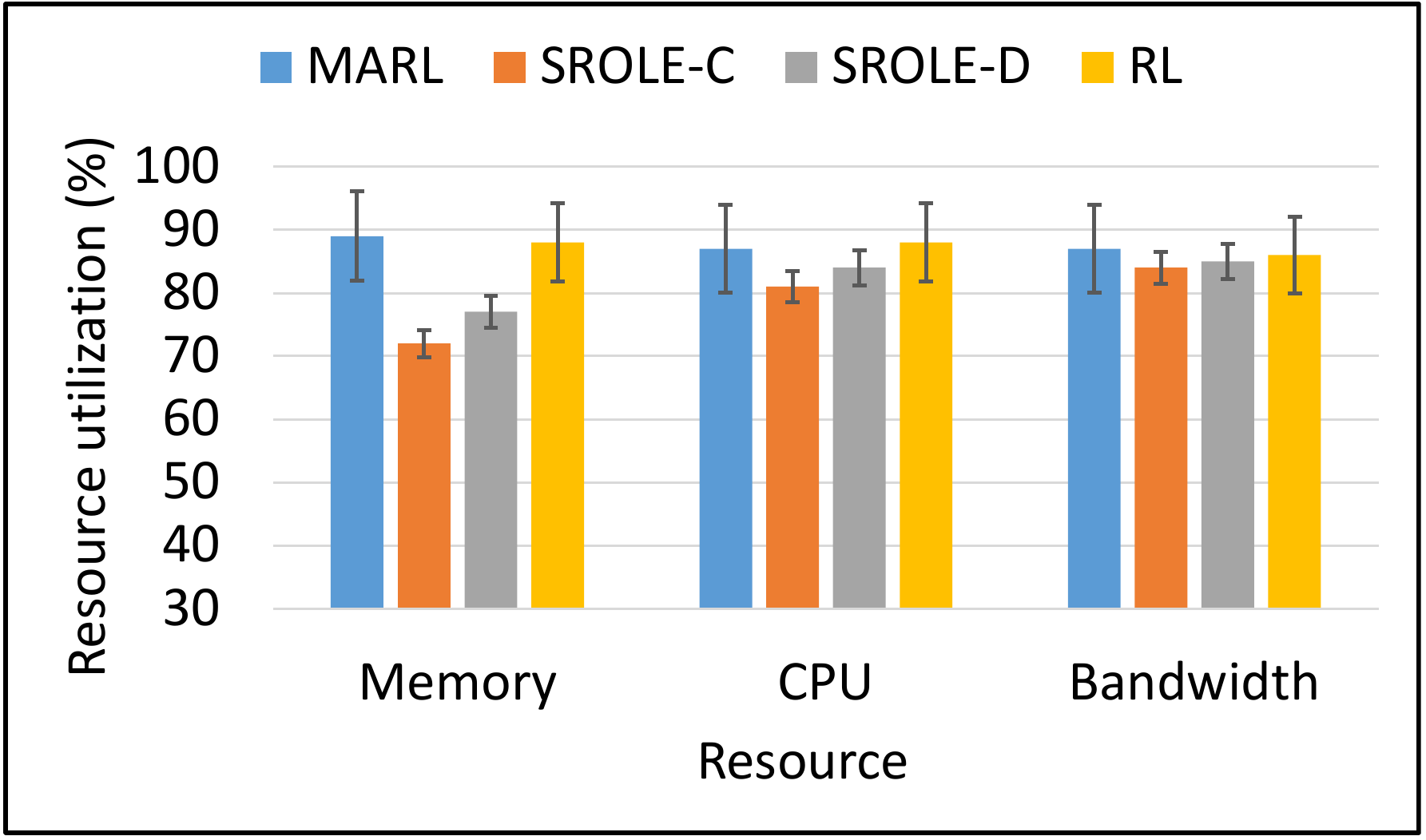} }}%
    \hfill
   \vspace{-0.08in}
   \caption{Resource utilization for different models from emulation.}%
    \label{fig:low2}\vspace{-0.2in}
\end{figure*}

\begin{figure*}[]
\centering
    \subfloat[VGG-16.\label{fig:low_mem2}]{{\includegraphics[width=0.32\linewidth,height=0.137\textheight]{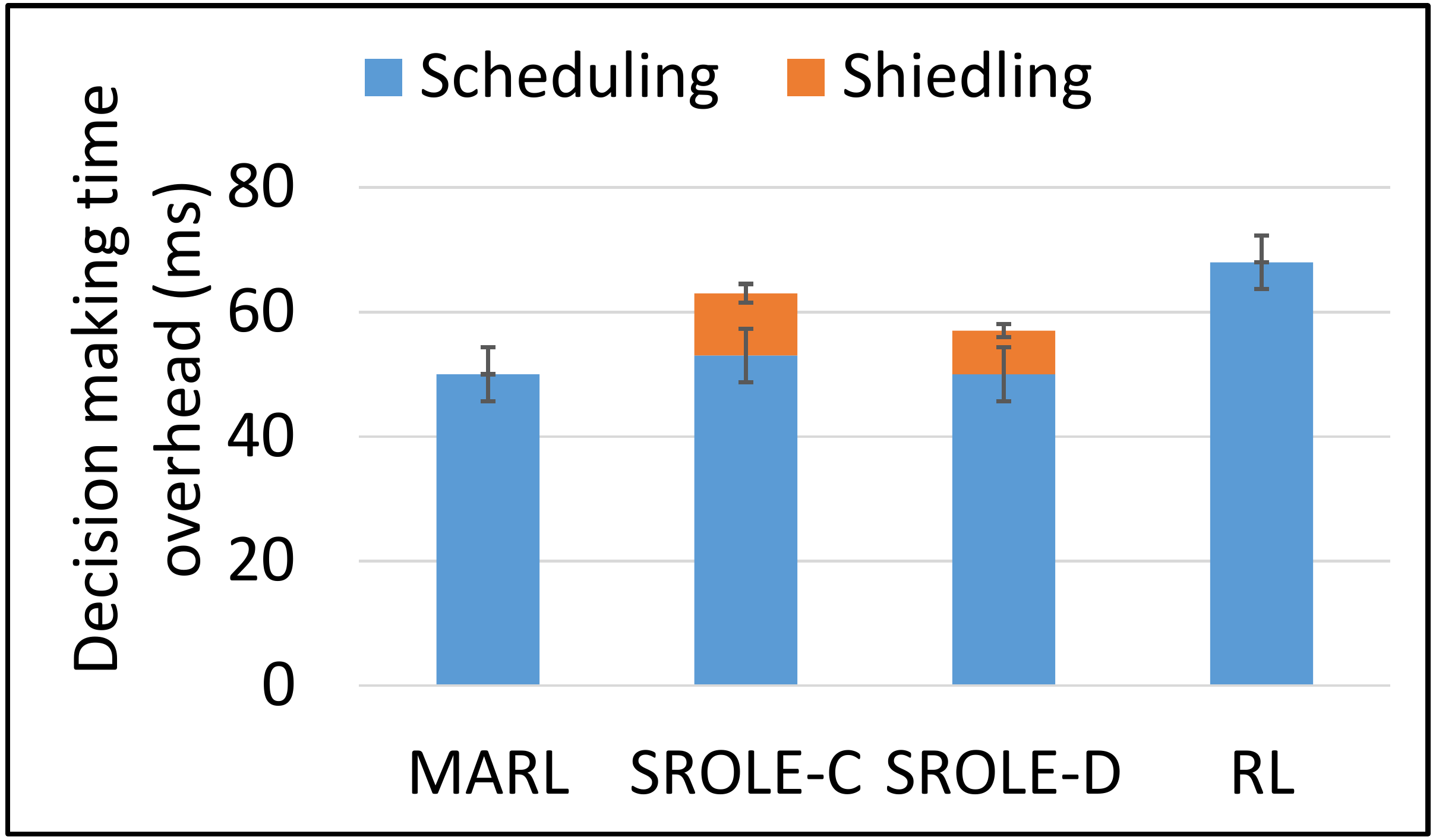} }}%
    \hfill
    \subfloat[GoogleNet. \label{fig:low_cpu2}]{{\includegraphics[width=0.32\linewidth,height=0.137\textheight]{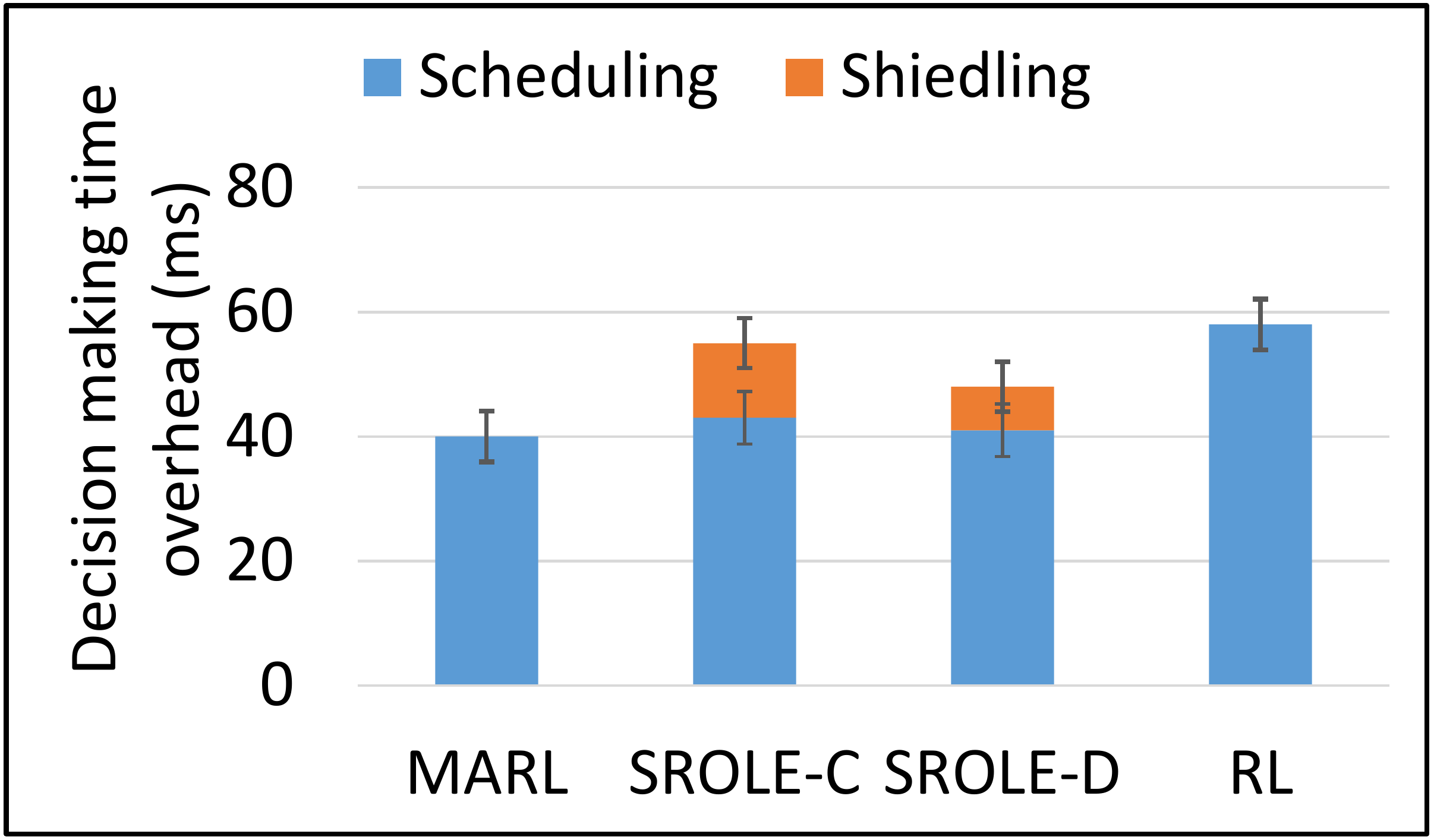} }}%
    \hfill
    \subfloat[RNN.\label{fig:low_bw2}]{{\includegraphics[width=0.32\linewidth,height=0.137\textheight]{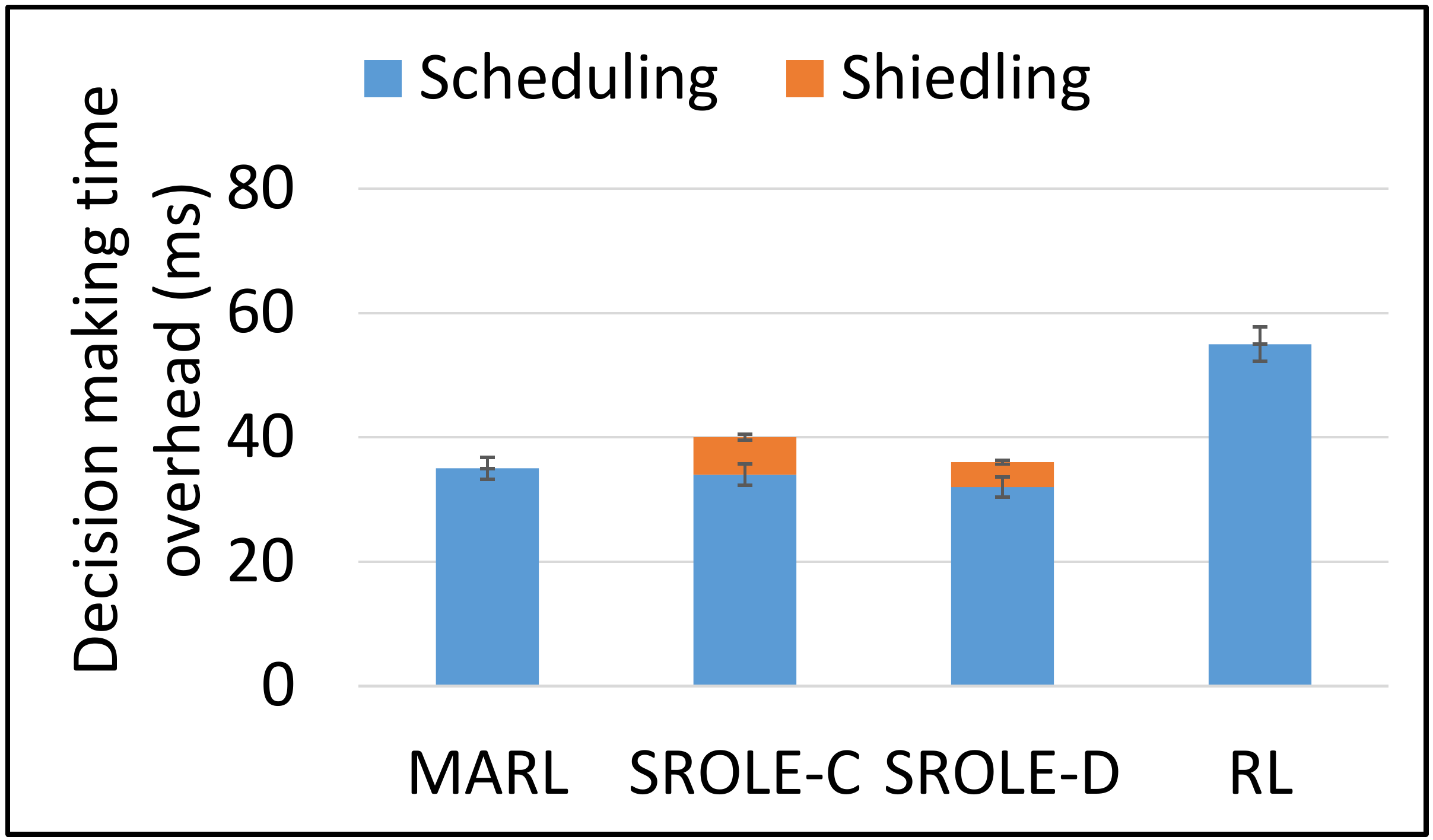} }}%
    \hfill
   \vspace{-0.08in}
   \caption{Computation overhead for different models from emulation.}%
    \label{fig:low4}\vspace{-0.2in}
\end{figure*}

\begin{figure*}[]
\centering
    \subfloat[VGG-16.\label{fig:low_mem3}]{{\includegraphics[width=0.32\linewidth,height=0.137\textheight]{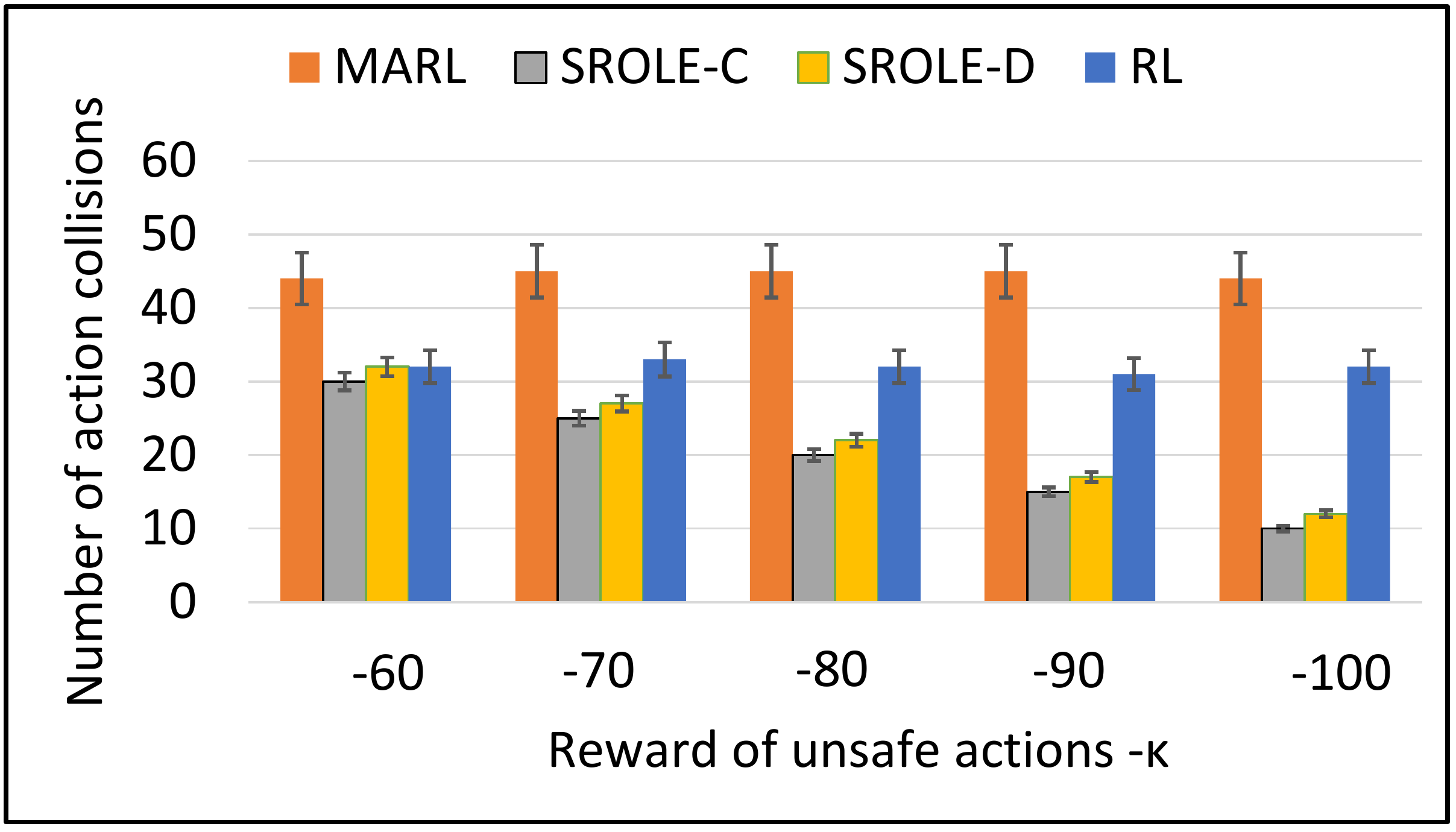} }}%
    \hfill
    \subfloat[GoogleNet. \label{fig:low_cpu3}]{{\includegraphics[width=0.32\linewidth,height=0.137\textheight]{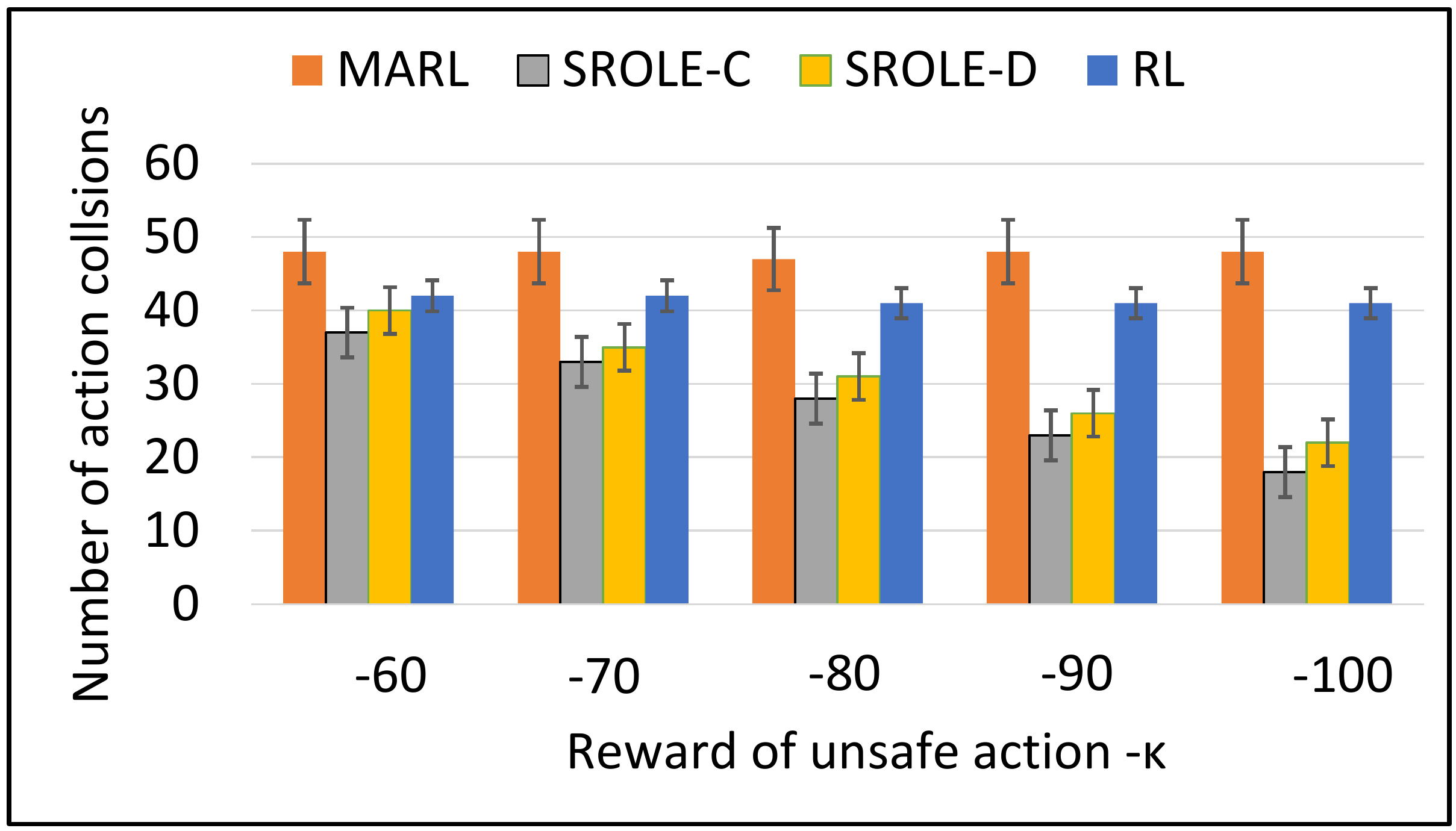} }}%
    \hfill
    \subfloat[RNN.\label{fig:low_bw3}]{{\includegraphics[width=0.32\linewidth,height=0.137\textheight]{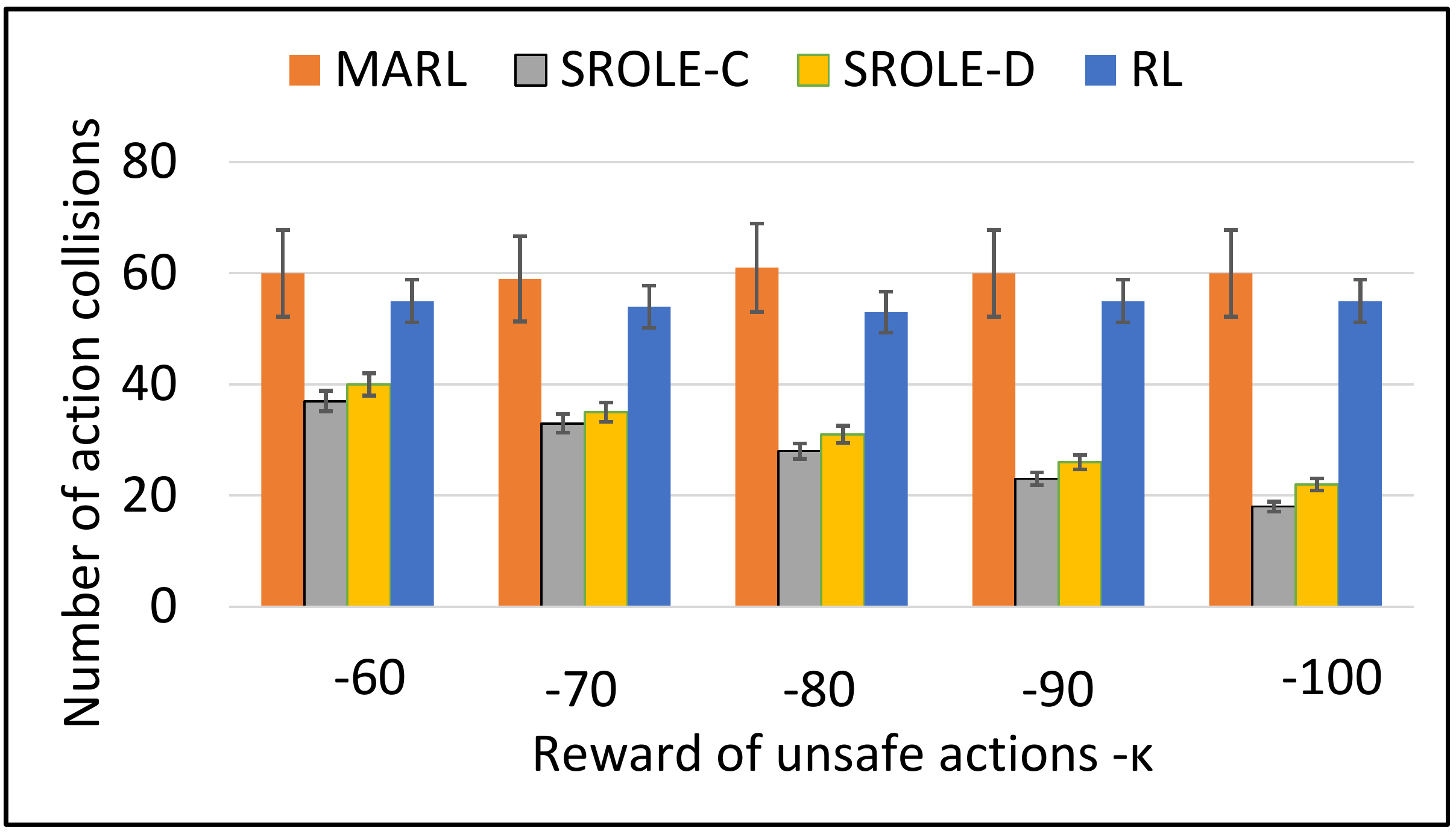} }}%
    \hfill
  \vspace{-0.08in}
   \caption{The number of action collisions for different models from emulation.}%
    \label{fig:low3}\vspace{-0.2in}
\end{figure*}

\vspace{-0.05in}
\noindent{\textbf{Job completion time.}}  Figures~\ref{fig:low_mem}, \ref{fig:low_cpu}, and \ref{fig:low_bw} show the job completion time versus the number of edges for training the VGG-16, GoogleNet, and RNN models, respectively. MARL and RL have similar job completion times, which indicates that the performances of their job schedules are similar. The results imply that MARL still can achieve comparable performance as RL though MARL does not have global information for job scheduling. Both centralized and decentralized shielding methods (SROLE-C and SROLE-D) perform better than RL and MARL without shielding because shielding reduces the number of unsafe actions, i.e., overloading individual devices. As a result, it reduces the job completion time.  For VGG-16, GoogleNet, and RNN, SROLE-D shows 36-45\%, 35-43\%, and 33-44\% reduction in job completion time than MARL or RL without shielding. SROLE-D performs 8-13\% less than SROLE-C for all three models because the action collisions are checked by multiple shields instead of one, adding extra communication time among the neighboring shields during the DNN training. As a result, SROLE-C saves job completion time by 49-56\% for VGG-16, 48-59\% for GoogleNet, 47-56\% for RNN, respectively in comparison with RL and MARL without shielding. Thus, the shielding can be conducted more efficiently because of observing the resource state of all the edges together. From all the figures, we see that as the number of edges increases, the job completion time increases. This happens because more clusters lead to more time in transferring the model parameters from the clusters to the parameter server for synchronization of the model. Also, the figures show the results do not vary greatly and keep relatively stable.\looseness=-1


\begin{figure*}[!t]
\centering
    \subfloat[VGG-16.\label{fig:low_memr}]{{\includegraphics[width=0.32\linewidth,height=0.137\textheight]{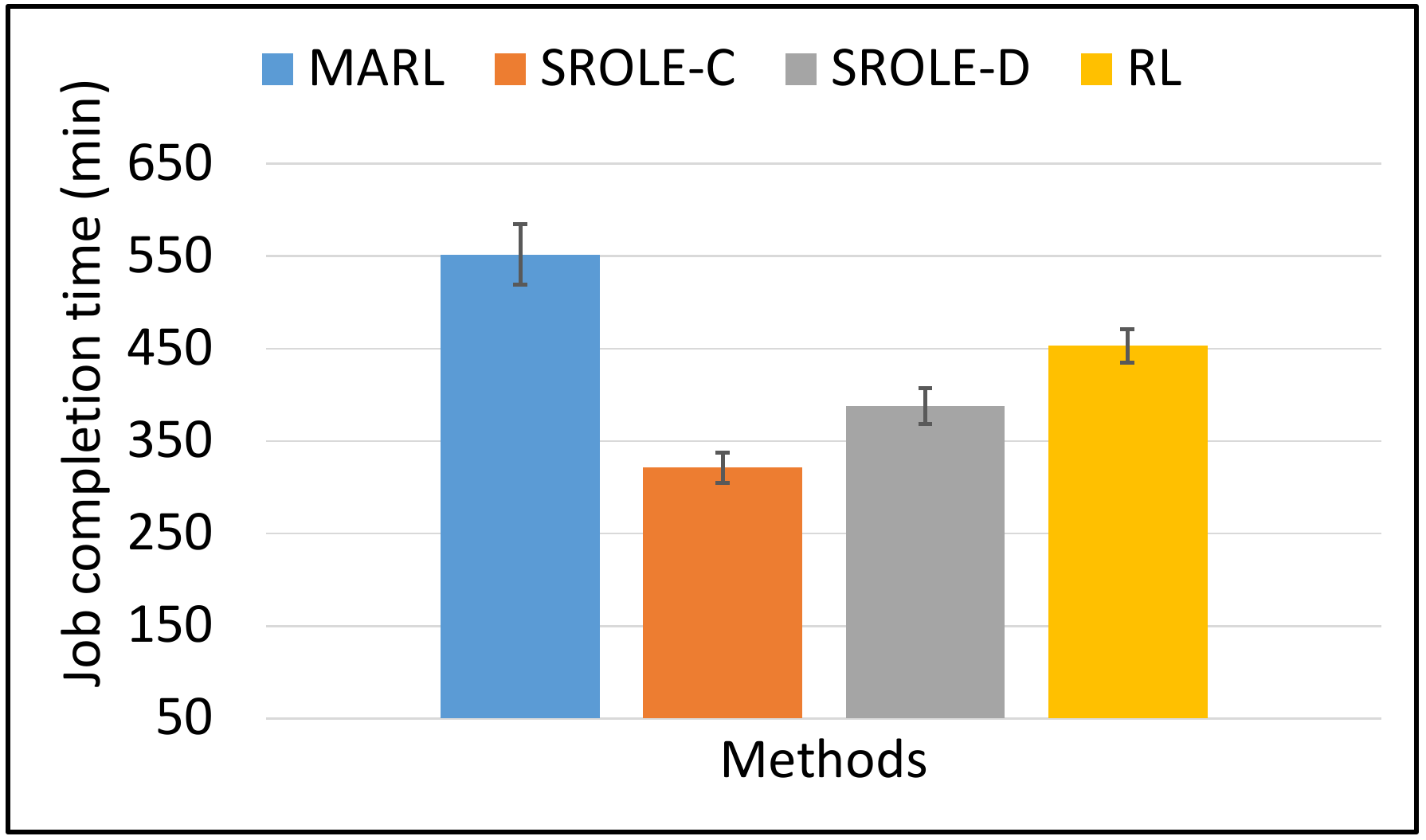} }}%
    \hfill
    \subfloat[GoogleNet. \label{fig:low_cpur}]{{\includegraphics[width=0.32\linewidth,height=0.137\textheight]{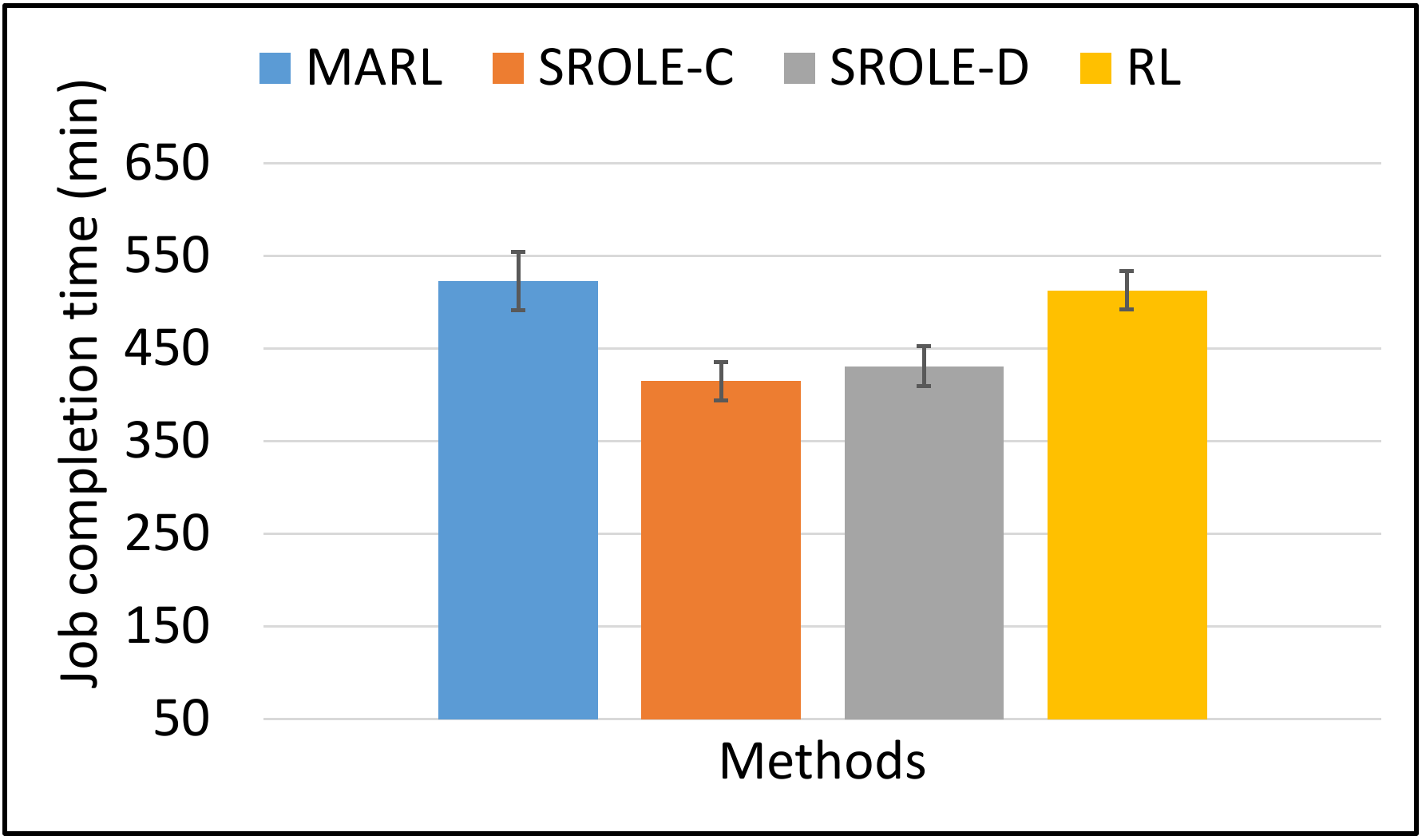} }}%
    \hfill
    \subfloat[RNN.\label{fig:low_bwr}]{{\includegraphics[width=0.32\linewidth,height=0.137\textheight]{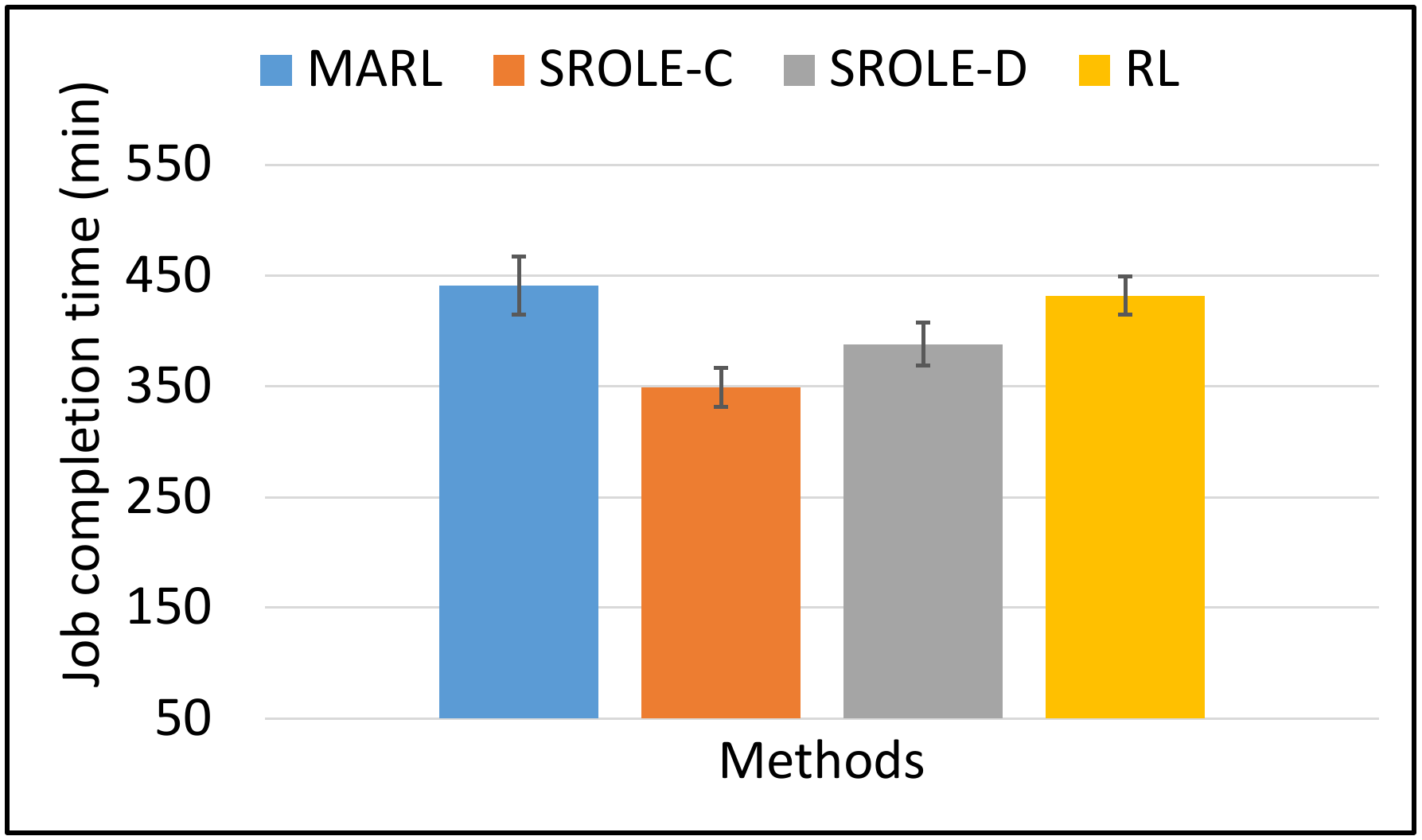} }}%
    \hfill
   \vspace{-0.08in}
   \caption{Job completion time for different models from a real-device network.}%
    \label{fig:lowr}\vspace{-0.2in}
\end{figure*}

\begin{figure*}[]
\centering
    \subfloat[VGG-16.\label{fig:low_mem1r}]{{\includegraphics[width=0.32\linewidth,height=0.137\textheight]{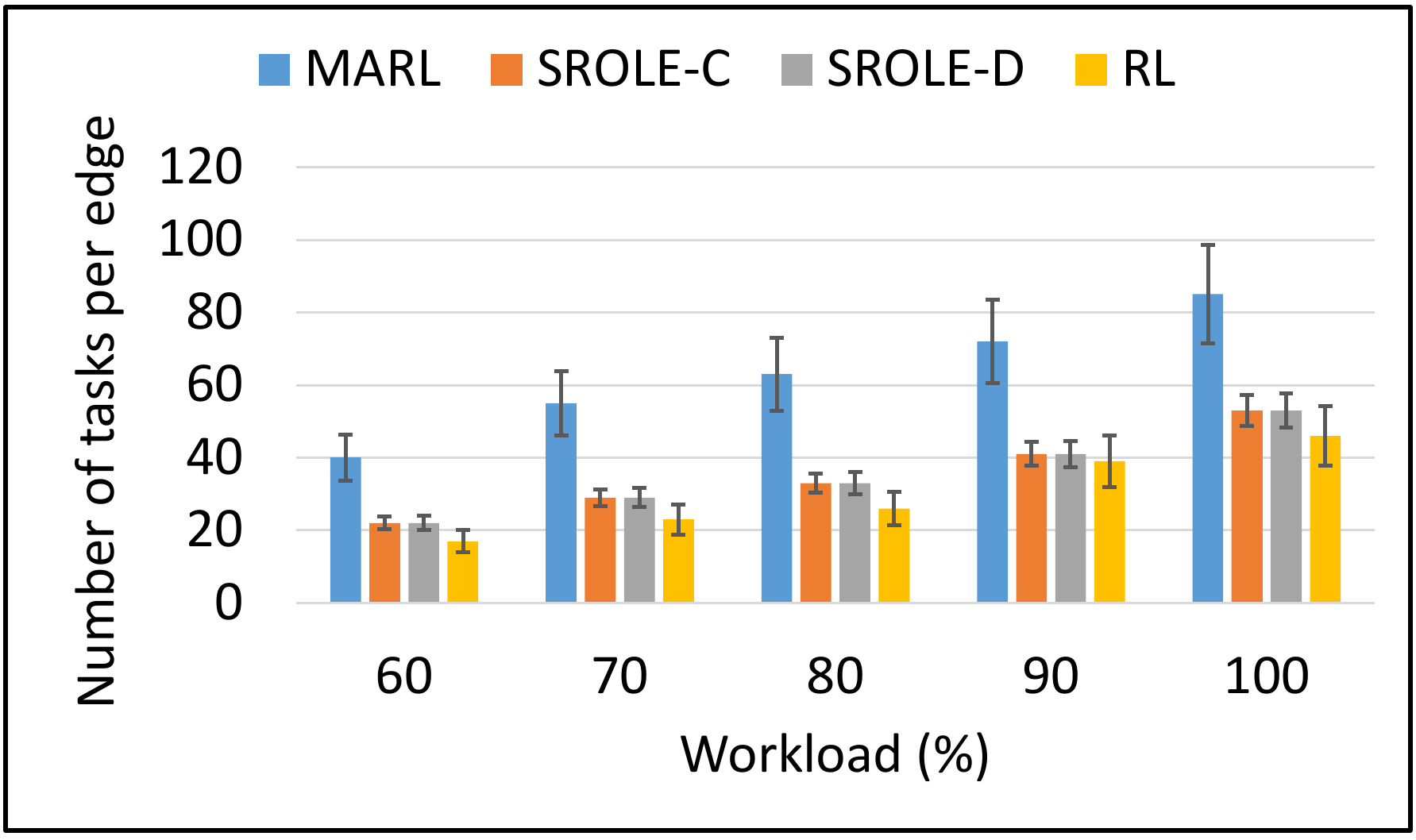} }}%
    \hfill
    \subfloat[GoogleNet. \label{fig:low_cpu1r}]{{\includegraphics[width=0.32\linewidth,height=0.137\textheight]{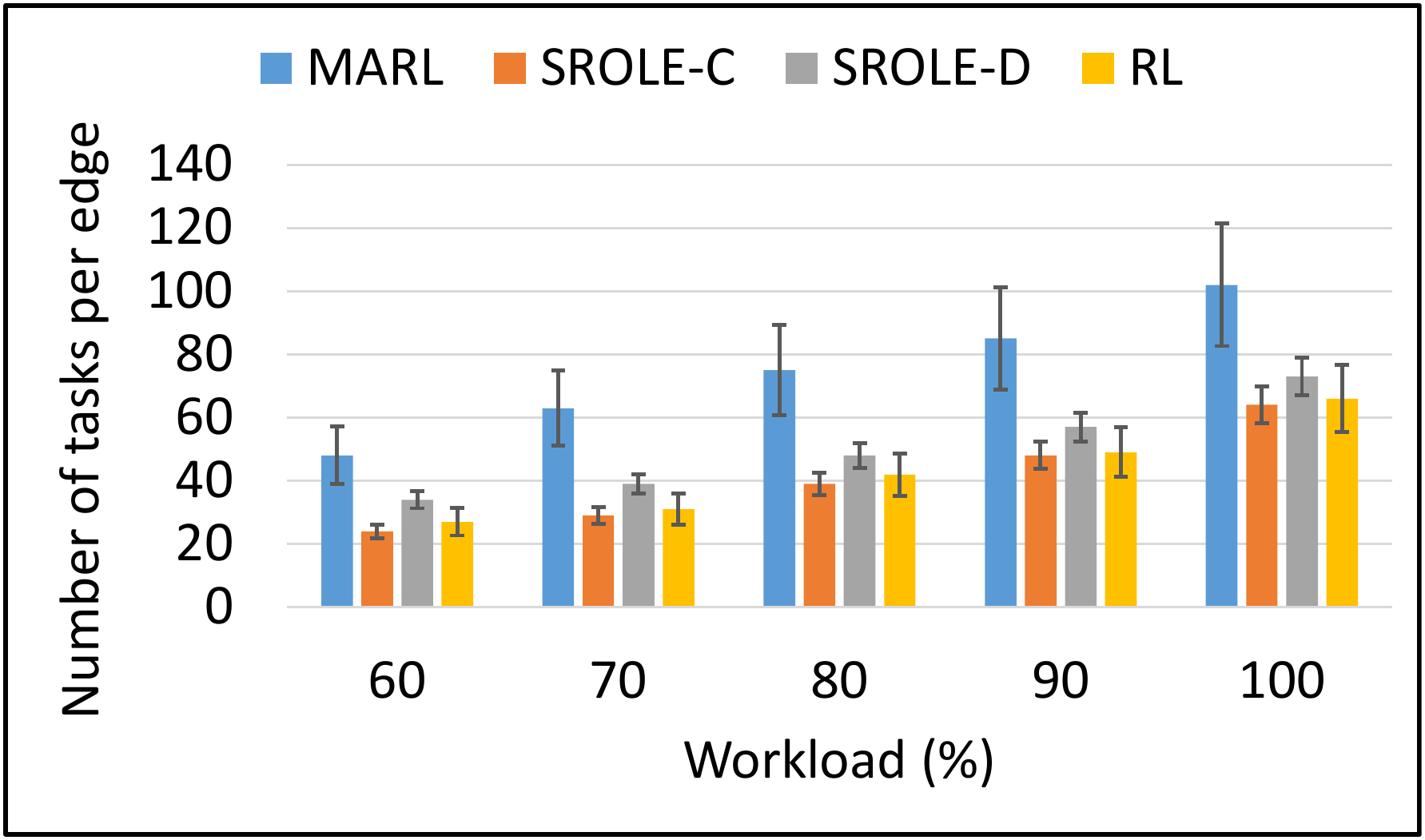} }}%
    \hfill
    \subfloat[RNN.\label{fig:low_bw1r}]{{\includegraphics[width=0.32\linewidth,height=0.137\textheight]{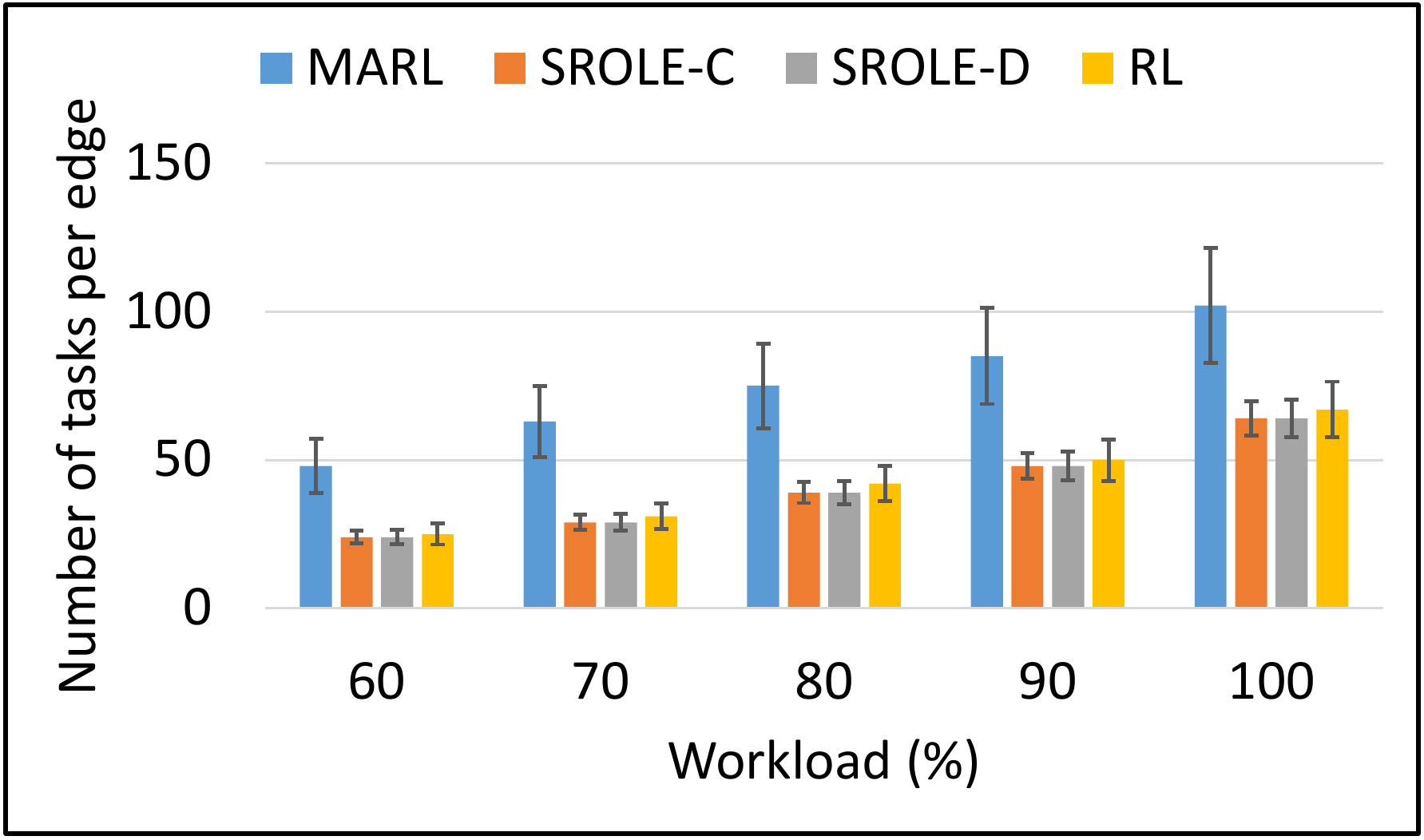} }}%
    \hfill
  \vspace{-0.08in}
   \caption{The number of tasks per device for different models from a real-device network.}%
    \label{fig:low1r}\vspace{-0.2in}
\end{figure*}
\begin{figure*}[]
\centering
    \subfloat[VGG-16.\label{fig:low_mem4r}]{{\includegraphics[width=0.32\linewidth,height=0.137\textheight]{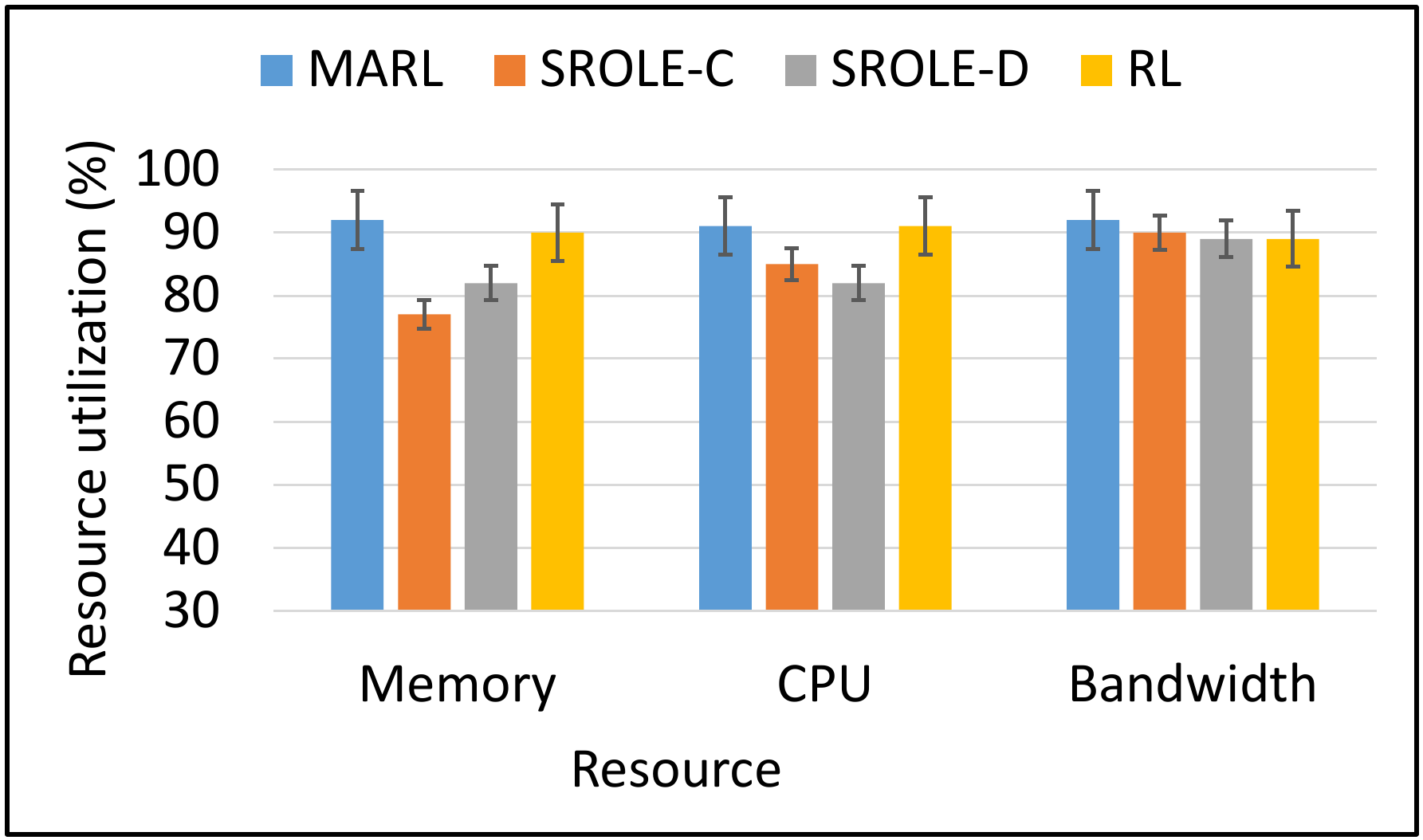} }}%
    \hfill
    \subfloat[GoogleNet. \label{fig:low_cpu4r}]{{\includegraphics[width=0.32\linewidth,height=0.137\textheight]{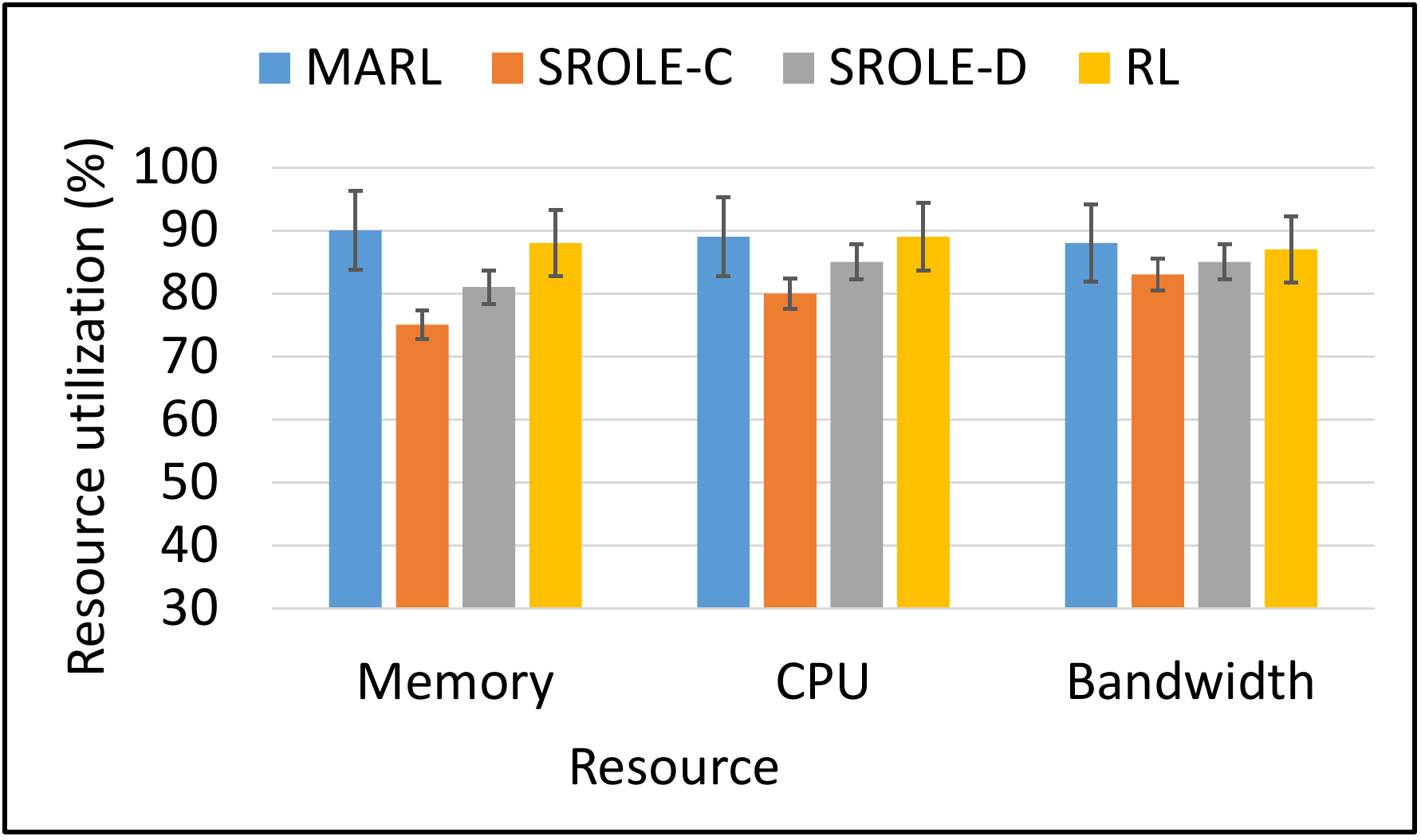} }}%
    \hfill
    \subfloat[RNN.\label{fig:low_bw4r}]{{\includegraphics[width=0.32\linewidth,height=0.137\textheight]{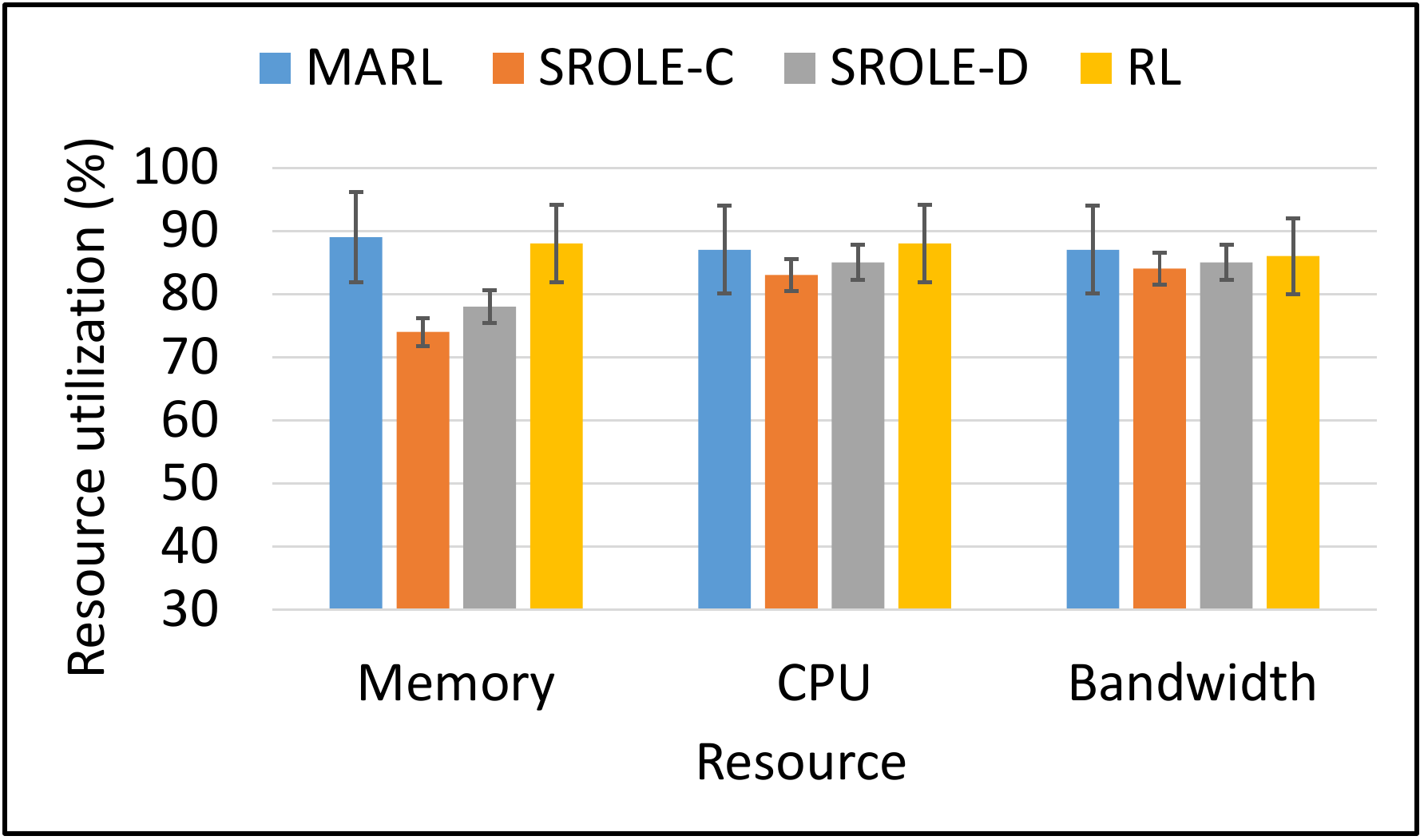} }}%
    \hfill
   \vspace{-0.08in}
   \caption{Resource utilization for different models from a real-device network.}%
    \label{fig:low2r}\vspace{-0.2in}
\end{figure*}

\vspace{-0.05in}
\noindent{\textbf{The number of tasks per edge node.}} Figures~\ref{fig:low_mem1}, \ref{fig:low_cpu1}, and \ref{fig:low_bw1} show the number of tasks per node versus different workloads for training the VGG-16, GoogleNet, and RNN models, respectively in the scenario with 25 edges. We plot the median (denoted with different colors) along with the minimum and the maximum number of tasks (denoted with black bars).   
\ts{SROLE-D shows 42-56\%, 46-61\%, and 41-56\% reduction in the median number of assigned tasks per device for VGG-16, GoogleNet, and RNN compared to MARL or RL without shielding, respectively. However, the SROLE-C outperforms the SROLE-D by 2-11\%. As a result, SROLE-C generates a reduction in the median number of assigned tasks per device by 49-56\% for VGG-16, 48-59\% for GoogleNet, 47-56\% for RNN respectively in comparison with RL and MARL without shielding.} From the figure, we also observe that both the SROLE-D and SROLE-C methods show less variance than both MARL and RL without shielding. Both SROLE-C and SROLE-D perform better than other methods because shielding reduces the number of unsafe actions, i.e., overloading on individual devices, and thus distributes the tasks among devices in a more balanced manner. However, SROLE-D performs less than SROLE-C shielding because of taking a higher number of unsafe actions. This phenomenon happens because of not knowing the cluster more completely.\looseness=-1


\vspace{-0.05in}
\noindent{\textbf{Resource utilization.}} Figures~\ref{fig:low_mem4}, \ref{fig:low_cpu4}, and \ref{fig:low_bw4} show the resource utilization of each type of resources for training the VGG-16, GoogleNet, and RNN models, respectively in the scenario of total 25 edges. We plot the median along with the minimum and maximum resource utilizations as error bars.   
\ts{ For VGG-16, GoogleNet, and RNN, SROLE-D shows 12-19\%, 11-17\%, and 11-15\% reduction in the median resource utilization compared to MARL or RL without shielding. However, the SROLE-C outperforms the SROLE-D by 2-14\%. As a result, SROLE-C generates a reduction in the median resource utilization by 21-24\% \% for VGG-16, 48-59\% for GoogleNet, 22-29\%  for RNN, respectively in comparison with RL and MARL without shielding.} From the figure, we also observe that both the SROLE-D and SROLE-C show less variance than both the MARL and RL without shielding in terms of resource utilization. Both centralized and decentralized shielding methods perform better than other methods because shielding avoids overloading individual devices. However, SROLE-D performs less than SROLE-C because of taking a higher number of unsafe actions. This happens due to not having the complete knowledge of the cluster, which adds extra computation or communication for the involvement of multiple shields.


\vspace{-0.05in}
\noindent{\textbf{Average computation time overhead.}} Figures~\ref{fig:low_mem2}, \ref{fig:low_cpu2}, and \ref{fig:low_bw2} show the computation overhead for scheduling (blue bar) and shielding (orange bar) of different methods while training the VGG-16, GoogleNet, and RNN models, respectively. For all the models, the results for computation overhead (scheduling + shielding) are as follow: MARL$<$SROLE-D$<$SROLE-C$<$RL. RL needs the longest decision making time because only one node is responsible for scheduling all jobs in one cluster. MARL greatly reduces the decision making time of RL since MARL distributes the scheduling load among the edge nodes in the cluster by letting each edge node schedule its own job among its neighbors. Both RL and MARL do not have any shielding time as they do not have the shielding approach. SROLE-C and SROLE-D are based on MARL, and they have additional shielding components to detect action collisions, thus generating higher decision making time. MARL, SROLE-C and SROLE-D have the same scheduling time since they all use MARL for scheduling. SROLE-D generates 5-8\% less shielding time than SROLE-C for all the models. This is because SROLE-C relies on one shield in each cluster, so the shield needs to check all the actions, which needs a long time, and SROLE-D distributes the shielding overload among multiple shields, which expedites the shielding process and reduces both the shielding time and the decision making time. 




\vspace{-0.05in}
\noindent{\textbf{The number of action collisions.}} Figure~\ref{fig:low3} shows how the assigned reward of unsafe action impacts the number of unsafe actions during the training of all the DNN models.
SROLE-C performs 31-48\% better than the MARL or RL, while SROLE-D performs 27-39\% better than MARL or RL for all the three models. This is because the added shield(s) in SROLE-C and SROLE-D coordinate the edges to avoid unsafe actions, but MARL and RL do not have shields. The SROLE-C approach performs 4-7\% better than SROLE-D. This is because the global shield in SROLE-C can observe the global environment, compare all actions and suggest alternative actions accordingly to avoid unsafe actions globally. When multiple shields are responsible for sub-clusters in SROLE-D, the information collected by a shield for the boundary nodes may not cover all the unsafe actions, leading to unsafe actions. Though SROLE-C and SROLE-D use the shielding approach, they still produce certain unsafe actions. This is because the resource demands of tasks are time-varying and dynamic and sometimes cannot be accurately predicted, thus leading to the edge node overload. For all the three models, as the absolute value of the reward of an unsafe action increases, the number of unsafe actions during the whole training period in SROLE-C and SROLE-D decreases and this number in MARL and RL keeps constant. MARL and RL do not use this reward or shielding approach, so their performances are not affected by the reward value. A high penalty for the action collision will help SROLE-C and SROLE-D avoid more unsafe actions.

\begin{figure*}[!t]
\centering
    \subfloat[VGG-16.\label{fig:low_mem2r}]{{\includegraphics[width=0.32\linewidth,height=0.137\textheight]{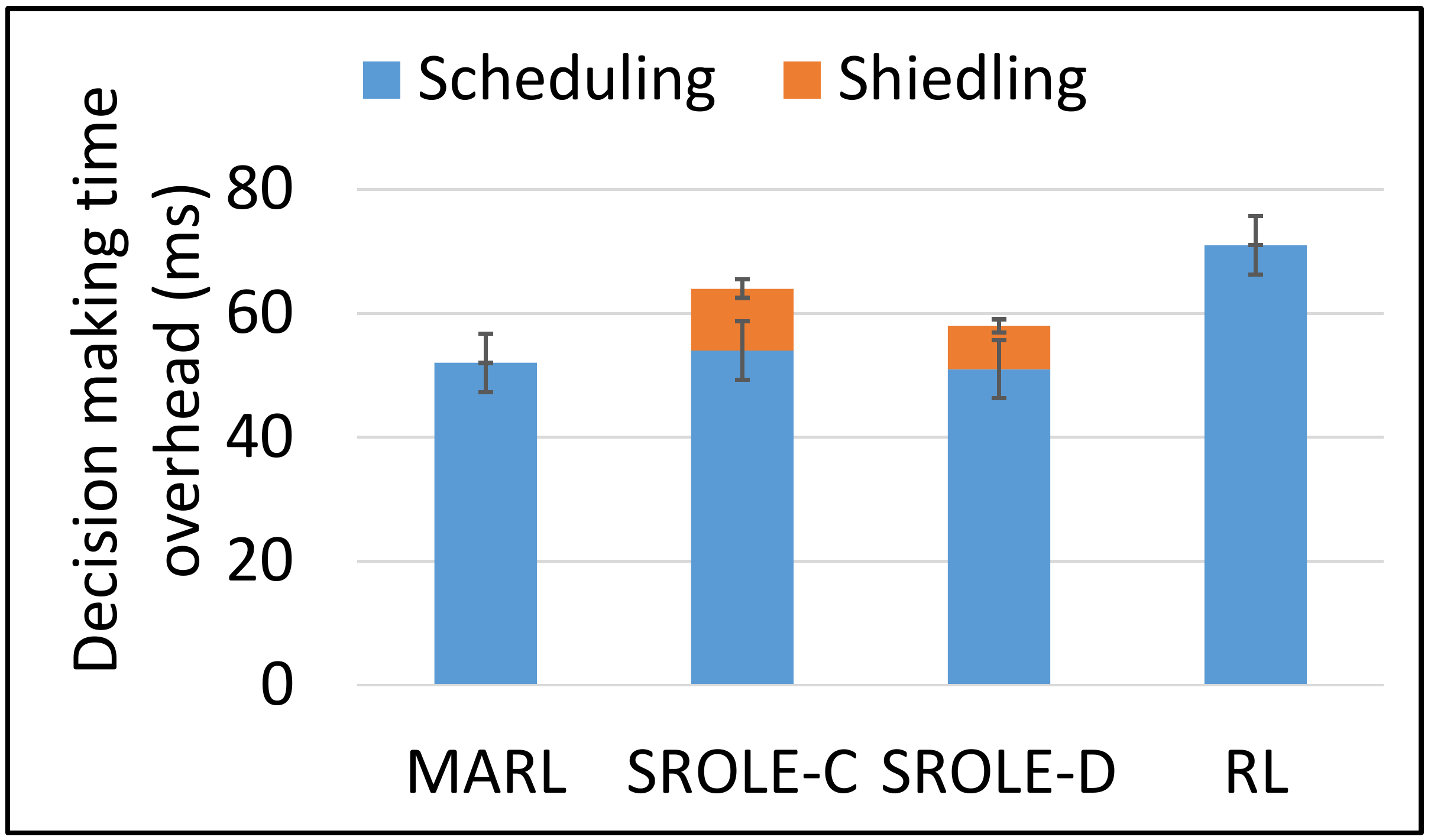} }}%
    \hfill
    \subfloat[GoogleNet. \label{fig:low_cpu2r}]{{\includegraphics[width=0.32\linewidth,height=0.137\textheight]{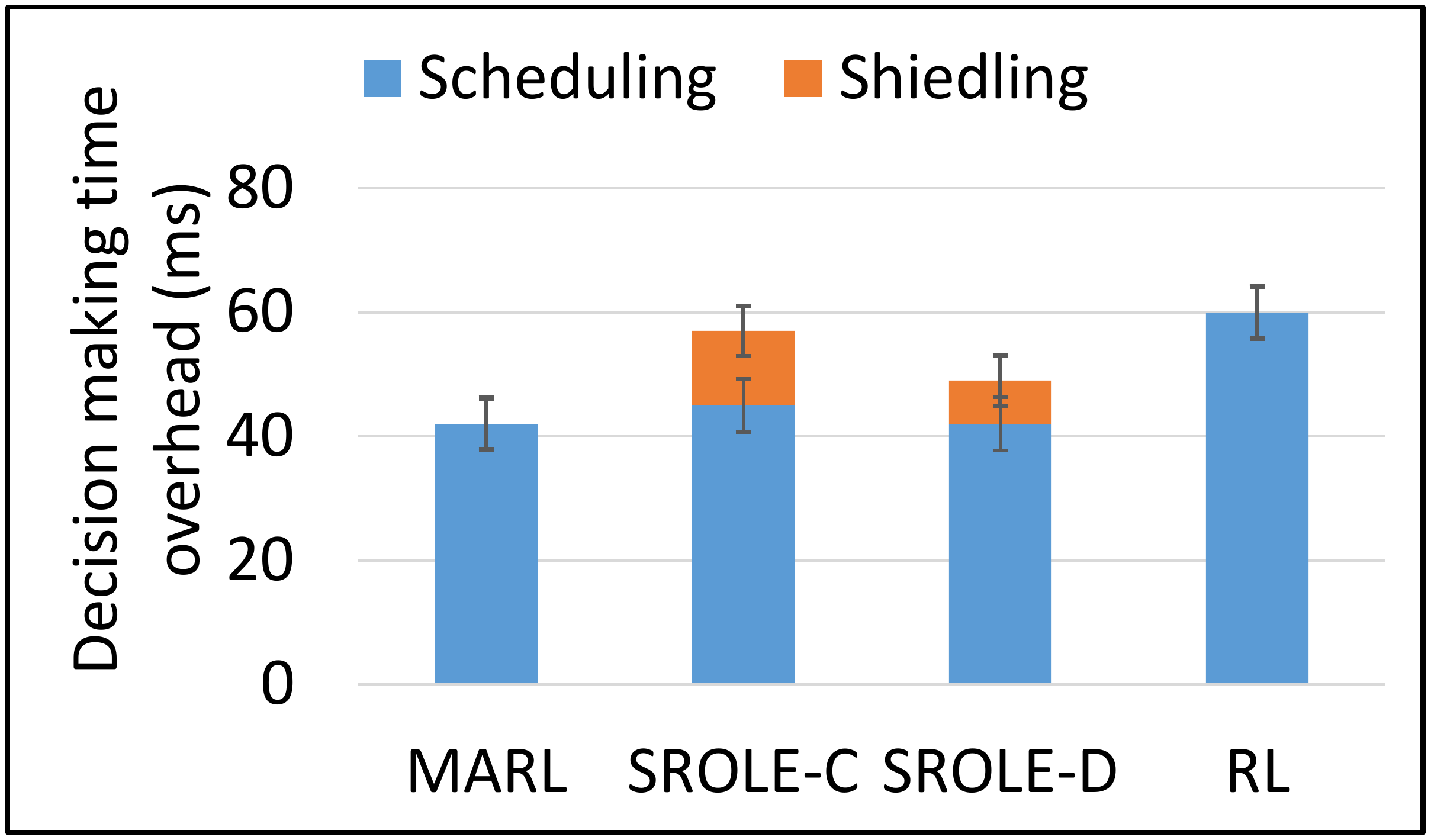} }}%
    \hfill
    \subfloat[RNN.\label{fig:low_bw2r}]{{\includegraphics[width=0.32\linewidth,height=0.137\textheight]{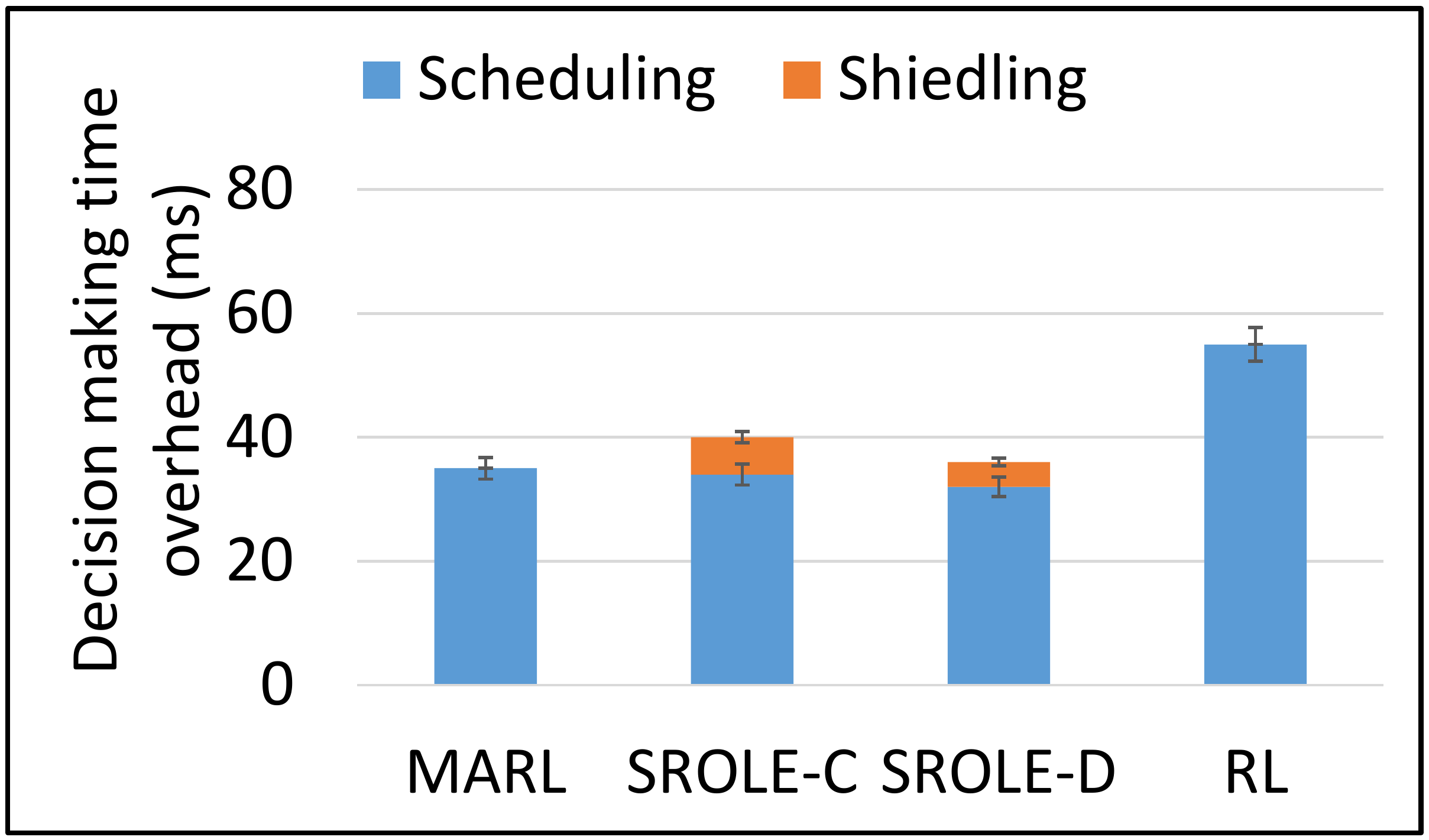} }}%
    \hfill
   \vspace{-0.08in}
   \caption{Computation overhead for different models from a real-device network.}%
    \label{fig:low4r}\vspace{-0.2in}
\end{figure*}

\begin{figure*}[!t]
\centering
    \subfloat[VGG-16.\label{fig:low_mem3r}]{{\includegraphics[width=0.32\linewidth,height=0.137\textheight]{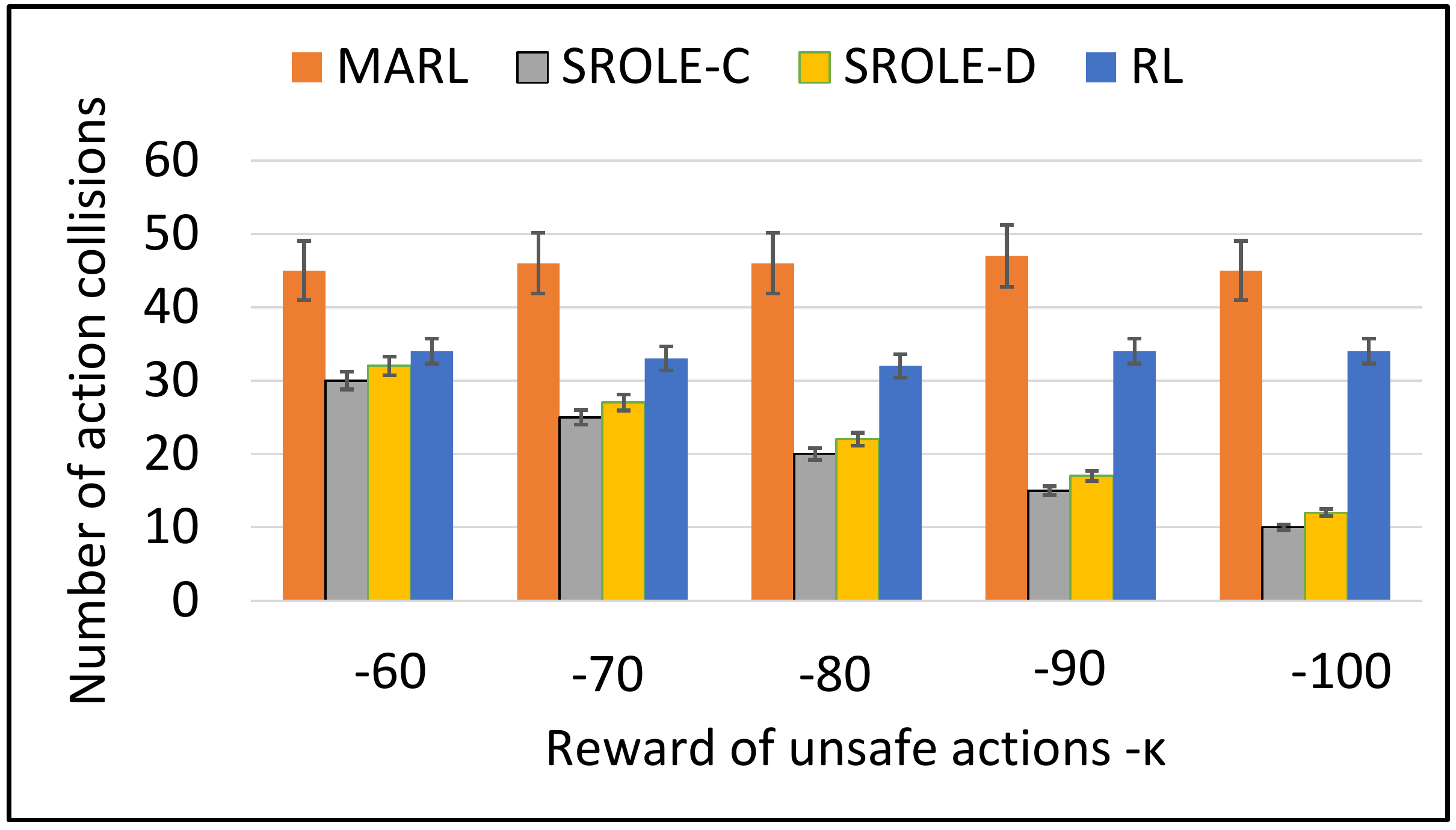} }}%
    \hfill
    \subfloat[GoogleNet. \label{fig:low_cpu3r}]{{\includegraphics[width=0.32\linewidth,height=0.137\textheight]{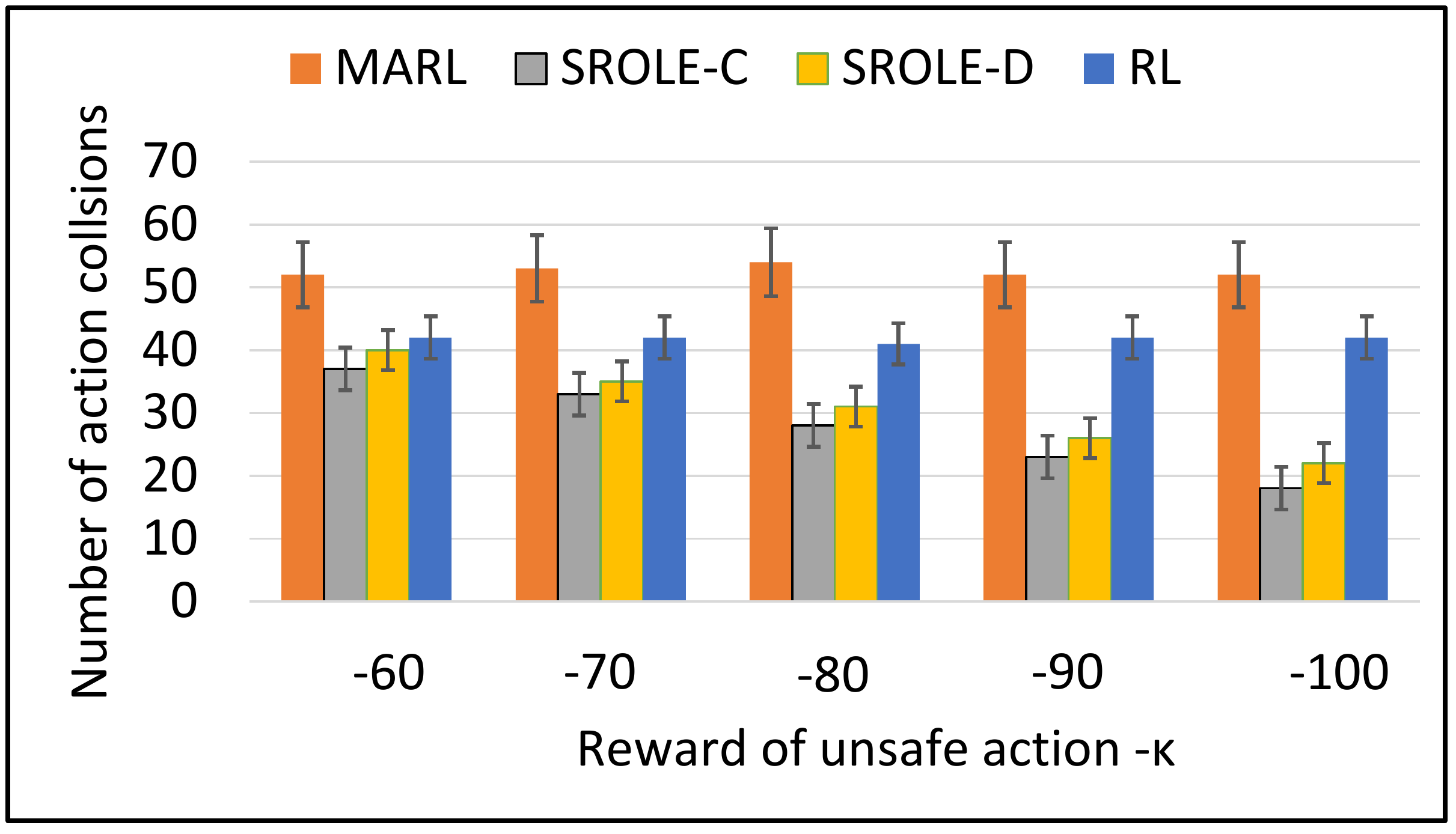} }}%
    \hfill
    \subfloat[RNN.\label{fig:low_bw3r}]{{\includegraphics[width=0.32\linewidth,height=0.137\textheight]{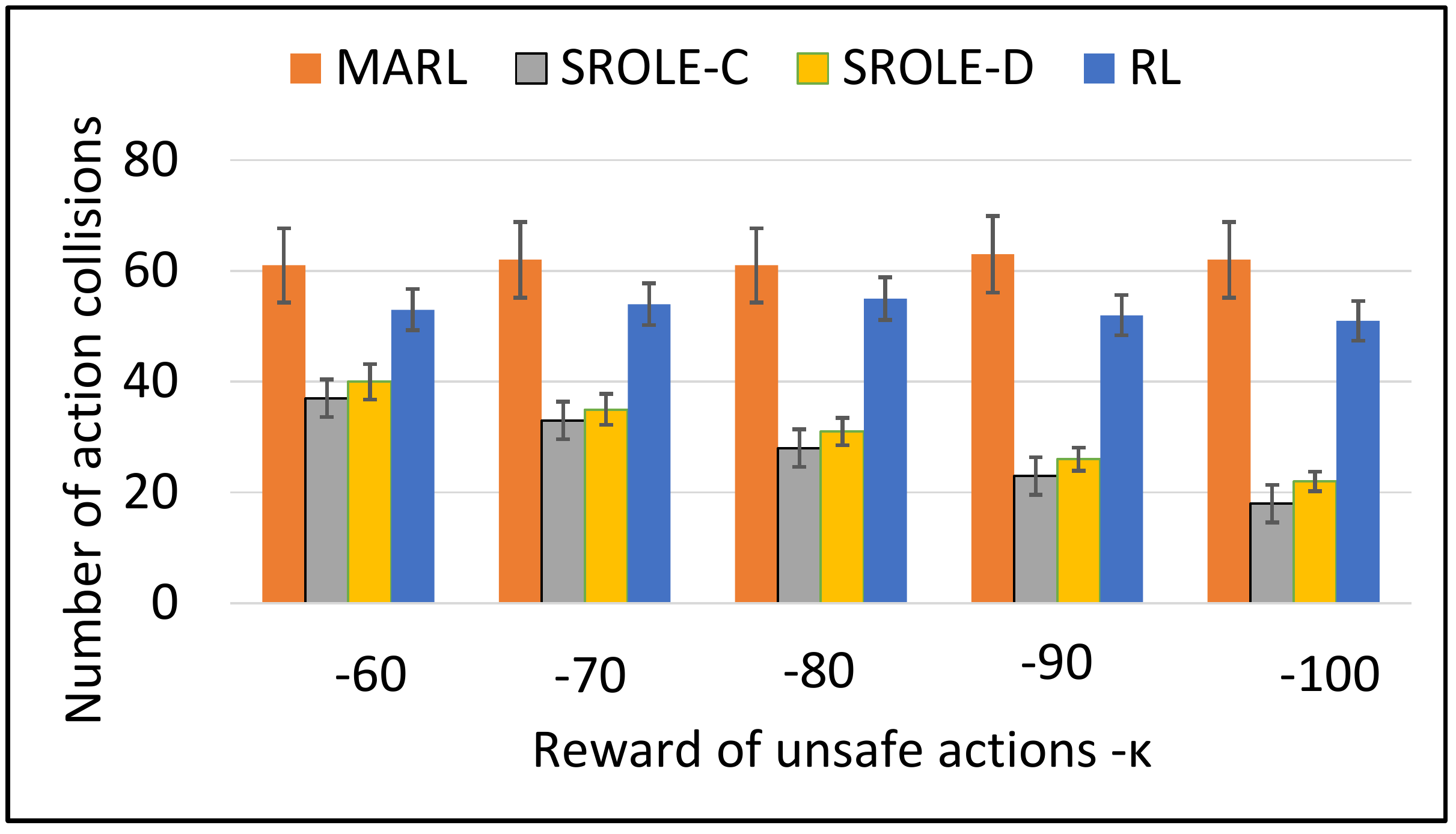} }}%
    \hfill
  \vspace{-0.08in}
   \caption{The number of action collisions for different models from a real-device network.}%
    \label{fig:low3r}\vspace{-0.2in}
\end{figure*}



\subsection{Experimental Results from a Real Device Network}\vspace{-0.05in} 
We formed the 10 edge nodes into a network and considered it as a single cluster, and then ran real experiments on the real-device network. From the real experiments, we observe similar performances of the different methods as in the container-based emulation due to the same reasons mentioned above. 
\noindent{\textbf{Job completion time.}}
Figure~\ref{fig:lowr} shows the the job completion time for training all models in the real-device cluster.
\ta{For these models, SROLE-D performs 32-39\% better than MARL or RL without shielding and SROLE-C performs 36-53\% better than MARL or RL. SROLE-D performs 4-7\% worse than SROLE-C because SROLE-D has additional communication operations between neighboring shields for shielding.}

\noindent{\textbf{The number of tasks per edge.}}
Figure~\ref{fig:low1r} shows the number of tasks per node for training all the models.
Comparing to MARL and RL without shielding, \ta{SROLE-D shows 28-45\% reduction and SROLE-C shows 39-52\% \ts{reduction in the median number of assigned tasks per device}. SROLE-D performs 7-11\% worse than SROLE-C because the shields in SROLE-D do not have global knowledge of the state of the resources in all edges.  Similar to the emulation experiments, the variances of SROLE-D and SROLE-C are lower than those of MARL and RL without shielding for all the models.}

\noindent{\textbf{Resource utilization.}}
Figure~\ref{fig:low2r} shows the resource utilization of each type of resources for training the three models.
\ta{SROLE-D shows 18-23\% reduction and SROLE-C shows 21-28\% \ts{reduction in the median in the median of resource utilization}. SROLE-D performs 3-5\% worse than SROLE-C because the shields in SROLE-D do not have global knowledge of the state of the resources in all edges.  Similar to the emulation experiments, the variances of SROLE-D and SROLE-C are lower than those of MARL and RL without shielding for all models.}\looseness=-1


\noindent{\textbf{Average computation time overhead.}}
Figure~\ref{fig:low4r} shows the computation overhead for scheduling and shielding of different methods while training all the models.
For the shielding time, SROLE-D performs 4-7\% better than SROLE-C for real-devices.\looseness=-1

\noindent{\textbf{The number of action collisions.}}
Figure~\ref{fig:low3r} shows how the assigned reward of unsafe action impacts the number of unsafe actions during training the DNN models.
\ta{SROLE-D shows 27-42\% reduction and SROLE-C shows 29-46\% \ts{reduction in the median in the median of resource utilization}. SROLE-D performs 2-6\% worse than SROLE-C because the shields in SROLE-D do not have global knowledge of the state of the resources in all edges.}

\section{Conclusion}
\vspace{-.05in}
\label{sec:concl}
Fully distributed DNN training on edges utilizing concurrent model and data parallelism is a promising way to increase the scalability of DNN model training at resource-constrained edges. However, relying on one edge node to use RL to schedule the model partitions (i.e., distributing the partitions of a large DL model to a set of edges to minimize the training time) among edge nodes is not scalable. In this paper, we propose a multi-agent RL system that enables each edge node to schedule its own jobs. To ensure such distributed job scheduling method will not overload an edge node, we propose using shielding that observes the actions decided by all edge nodes to avoid overloading edge nodes. Relying on one shield is not scalable. Thus, we further propose a decentralized shielding method that relies on multiple shields to conduct shielding in a distributed manner. Our experiments show that our shielding method performs 59\% better than multi-agent RL in training time with 29\% less median resource utilization of an edge device, and also the multi-agent RL method achieves similar job completion time performance as the centralized RL method. In the future, we will explore using formal method approaches to provide a guarantee of action collision avoidance and explore a method to avoid action collisions caused by decentralized shielding.\looseness=-1

\section*{Acknowledgements} \label {sec:Acknowledgements}
\vspace{-.05in}
This research was supported in part by U.S. NSF grants NSF-1827674, CCF-1822965, FHWA grant 693JJ31950016, Microsoft Research Faculty Fellowship 8300751, and the Commonwealth Cyber Initiative (CCI), an investment in the advancement of cyber research, innovation and workforce development. For more information about CCI, visit cyberinitiative.org. We also thank Ms. Ingy ElSayed-Aly and Dr. Lu Feng for helping us with our initial understanding of RL shielding.

\footnotesize{
\bibliographystyle{IEEEtran}
\bibliography{myBib}}
\end{document}